\documentclass{aa}
\usepackage{natbib}
\bibpunct{(}{)}{;}{a}{}{,} 

\usepackage{graphicx}
\usepackage{caption}
\usepackage{subcaption}
\usepackage{amssymb}
\usepackage{comment}
\usepackage{txfonts}
\usepackage[dvipsnames]{xcolor}
\usepackage[colorlinks=true,citecolor=RoyalBlue, linkcolor=Emerald]{hyperref}

\usepackage[normalem]{ulem}

\begin{document}

   \title{Super-Earth formation in systems with cold giants}

   \author{Claudia Danti\inst{1}
          \and
          Michiel Lambrechts\inst{1}
          \and
          Sebastian Lorek\inst{1}
          }

   \institute{Center for Star and Planet Formation, Globe Insitute, Øster Voldgade 5, 1350 Copenhagen, Denmark\\
              \email{claudia.danti@sund.ku.dk}\\
              \email{danticlaudia@gmail.com}
             }

   \date{}

  \abstract
   {

Around our Sun, terrestrial planets did not grow beyond Earth in mass, while super-Earths are found to orbit approximately every other solar-like star.
It remains unclear what divides these super-Earth systems from those that form terrestrial planets, and what role wide-orbit gas giants play in this process.
Here, we show that the key uncertainty is the degree of viscous heating in the inner disc, which regulates the pebble accretion efficiency.
In this parameter study, we assume pebble sizes limited by fragmentation and radial drift. The initial seed planetesimals for embryo growth are taken from the top of the streaming instability mass distribution. 
We then evaluate the important role of the pebble scale height and the assumed pebble fragmentation velocity. In systems with maximally efficient viscous heating, where all the accretion heating is deposited in the disc midplane, pebble accretion in the terrestrial region is suppressed. More realistic levels of viscous heating, at higher elevations, allow terrestrial embryo formation at Earth-like orbits. 
We also find that the role of the water iceline is minor, unless it is paired with extreme volatile loss and a change in the pebble fragmentation velocity.
Furthermore, we show that in systems with gas-giant formation, the role of mutual pebble filtering by outer pebble-accreting embryos is limited, unless some mechanism of delaying inner disc growth, such as viscous heating or the presence of an iceline, is simultaneously employed.
This latter point appears to be consistent with the fact that no strong suppression is seen in the occurrence rate of super-Earths in systems with known gas giants in wider orbits.
We conclude that the diversity in inner-disc systems may largely be driven by complex, and as of yet poorly understood, disc accretion physics inside the water ice line.}


   \maketitle
%

\section{Introduction}

Protoplanetary discs are the birth environment of planets.
During their evolution, dust grains collide and grow from well-coupled micrometre-sized particles, to pebbles in the mm to cm size range \citep{brauer_coagulation_2008,guttler_outcome_2010, zsom_outcome_2010}, which then start drifting inward towards the star \citep{adachi_gas_1976, weidenschilling_aerodynamics_1977,brauer_coagulation_2008}.
As pebbles grow in size, they settle down towards the disc midplane, increasing the local dust-to-gas ratio, where the streaming instability can trigger the formation of dense swarms of pebbles that undergo gravitational collapse, building planetesimals \citep{youdin_particle_2007, johansen_rapid_2007, bai_dynamics_2010}. Through this mechanism, it is possible to bypass the so-called metre-size barrier, the size barrier above which collisions among pebbles are entirely disruptive, and directly form km-size (and larger) planetesimals \citep{Schafer2017,johansen_formation_2015}.
The largest planetesimals can then act as planetary embryos and start the planetary growth process, through collisions with other kilometre-size planetesimals (``planetesimal accretion'') and/or by sweeping up inwards-drifting pebbles (``pebble accretion''). 

In the outer disc, core growth by accretion of planetesimals only can be slow, especially outside Jupiter-like orbital distances \citep{pollack_formation_1996, Rafikov_2004, Levison_2010}, making it incompatible with the rapid core formation timescales needed to form cold giants within a disc lifetime of few Myr. 
The pebble accretion scenario \citep{ormel_effect_2010, lambrechts_rapid_2012} provides a faster growth timescale, since the mm to cm size pebbles are efficiently captured by the planetary embryo, due to the dissipation of their kinetic energy by gas drag. 
Planetary embryos accrete pebbles until they reach the so-called pebble isolation mass, a mass sufficiently high to open a shallow gap in the disc, locally inverting the pressure gradient, and halting the drift of pebbles \citep{lambrechts_separating_2014, bitsch_pebble-isolation_2018}. After reaching the pebble isolation mass, the planet undergoes an envelope contraction phase, possibly followed by runaway gas accretion. In our Solar System this mechanism is a potential explanation for the dichotomy between gas and ice giants, with the latter not reaching isolation mass within the disc lifetime, thus never undergoing runaway gas accretion \citep{pollack_formation_1996,lambrechts_separating_2014, venturini_formation_2017, frelikh_formation_2017}.

In the inner disc, approximately within the water ice line, the role of pebble accretion is less clear. 
Closer to the star, embryo growth may be driven through planetesimal and embryo-embryo collisions if these are present with a high surface density, in a scenario reminiscent of classic gas-free terrestrial planet formation \citep{Chambers_1998, Mordasini_2012, Emsenhuber_2021}.
The ubiquitous presence of super-Earths and sub-Neptunes around nearly half Sun-like starts \citep{He_architectures_2021}, some with H/He envelopes \citep{Fulton_carlifornia_2017, Misener_2021}, argues that, at least in some systems, growth of inner disc embryos is completed before gas disc dissipation. 
This implies that these cores can accrete through pebble accretion and, when crossing Mars in mass, also type-I migrate through the disc \citep{Tanaka_2002}. 

Early N-body work, focused on planetsimal accretion only, showed that the high planetesimal surface densities required for super-Earth formation in a nominal gas disc lead to systems of planets migrating to the disc edge, with generally a poor match to the exoplanet population in mass and orbital parameters \citep{Ogihara_reassessment_2015}. 
Many works have explored this question further by addressing these shortcomings with 
discrete high-surface-density planetesimal rings \citep{morbidelli_contemporary_2022,batygin_formation_2023} and 
reduced migration in disc-wind sculpted discs \citep{Ogihara_suppression_2015,Ogihara_2018, Ogihara_2024}, 
or by including heating torques \citep{Broz_2021}. 
Pebble accretion can further aid the growth of inner-disc bodies, either by completing the growth of super-Earths/sub-Neptunes during the gas disc phase, or, in case of a lower pebble surface densities, stalling the embryos at Mars mass, which can subsequently form terrestrial planet-like systems through a giant impact phase \citep{lambrechts_formation_2019}. 

Recent studies have started to employ hybrid models of pebble and planetesimal accretion, both in the context of the formation of the Solar System \citep{ Levison_growing_gas_2015, Levison_growing_terr_2015, Lichtenberg_2021}
and exoplanetary systems \citep{bitsch_rocky_2019,izidoro_formation_2021, bitsch_giants_2023}, studying the efficiency of pebble and planetesimal accretion as concurrent processes. \citet{Levison_growing_terr_2015} invoked viscously stirred pebble accretion to circumvent the problem of forming too many super-Earths when the pebble flux is high enough to create gas giants, proposing that the hybrid model is able to reproduce both gas giants and suppress super-Earth formation. \citet{bitsch_giants_2023} were able to reproduce the conditional occurrence rates of cold Jupiters and inner sub-Neptunes, also matching the eccentricity distribution of giant exoplanets.

In summary, the magnitude of the pebble accretion contribution in the inner disc is uncertain, but if pebbles are invoked for the formation of planetesimals, it is difficult to imagine that they do not also play a role in embryo growth.

Observationally, it is found that cold giants do not suppress the growth of super-Earths, but rather either enhance it, or do not  affect it substantially.
\citet{rosenthal_california_2022} inferred a conditional probability of the occurrence rate of close-in small planets
when a Jupiter-like planet is present,
of $P(\mathrm{I|J}) = 32_{-24}^{+16}\%$ around FGKM stars, which does not differ significantly from the field occurrence of super-Earths of $P(I)\approx 30 \%$ \citep{zhu_about_2018}. More recent studies either agree on the absence of correlation in the joint occurrence rate \citep{bonomo_cold_2023} or on a slight enhancement of it \citep{van_zandt_2025}.
Further studies also analysed the metallicity of the host star, showing how the correlation might be positive for metal-rich stars and disappear for metal-poor stars \citep{bryan_friends_2024}.

In this work, we focus on systems that formed cold giants in wide orbits. 
This will allow us to explore, in general, the occurrence rate of super-Earths/sub-Neptunes in systems with cold giants, which has become an area of considerable interest in exoplanet occurrence studies. 
We define as a super-Earth any planet that exceeds Earth in mass ($1 \: M_{\oplus}<M \lesssim 20 \: M_{\oplus}$) and orbits in a closer orbit ($a_{\rm p}<1 \: \rm AU$). We also do not distinguish between super-Earths and sub-Neptunes as the final composition of the planet is not object of this study.

Previous theoretical works have investigated the super-Earths to cold giants correlation in different ways. \citet{chachan_small_2023} found that discs that are massive enough to nucleate heavy cores around 5 AU possess more than enough material able to drift inwards and build up inner planets, giving rise to the observed positive correlation
between inner planets and outer giants.
\citet{bitsch_giants_2023} performed N-body studies and identified the importance of gas accretion rates on the formation of systems of inner sub-Neptunes and outer gas giants, showing that less efficient envelope contraction rates allow for a more efficient formation of systems with inner sub-Neptunes and outer gas giants.
Also classical planetsimal accretiom models can reproduce super-Earths and cold Jupiters positive correlation, even if in a weaker fashion than suggested by observations \citep{Schlecker_2021}.

Our paper is structured as follows.
Section \ref{sect:model} is devoted to construct the theoretical basis of our model. Section \ref{subsect:gas_disc} describes how we derive the disc aspect ratio (and gas surface density) of the disc, that critically depends on the height at which viscous heat dissipation takes place.  
Section \ref{sec:pebble_disc} describes the radial pebble flux, with particular focus on the role of fragmentation- and drift-limited pebbles.
We present the results of our analysis in Section \ref{sect:results}, which is organised into four different subsections aimed at exploring the key uncertainties that this study focuses on.
In Section \ref{subsect:role_vfrag}, we investigate the role of the uncertain pebbles sizes due to poorly understood fragmentation velocities that depend on material and monomer size assumptions.
Section \ref{subsect:role_visc} discusses the effects of accretion heating in the inner disc on inner-planet formation efficiency.
The effects of pebble filtering due to outer giant planets up to isolation mass are presented in Section \ref{subsect:role_multiple_giants}, while Section \ref{sec:leaking_dust} analyses how pebbles leaking through the planet's gap impact the inner embryo's growth.
Section \ref{subsect:role_iceline} focuses on studying the effects of the iceline on inner-planet formation.
Finally, Section \ref{sec:discussion} is devoted to discussion, model assumptions, and avenues for further work. 
Our conclusions are summarised in Section \ref{sec:conclusion}.

\section{Model}
\label{sect:model}

We conduct our study using a 1D model to describe the protoplanetary disc and planetary growth.
In the following sections, we describe the analytical expressions used for the gas and pebble disc (Sections \ref{subsect:gas_disc}, \ref{sec:pebble_disc}), as well as the pebble accretion equations (Section \ref{subsec:plangrowth}).

\subsection{Gas disc model}
\label{subsect:gas_disc}
The disc structure is a fundamental choice in our model, as it regulates processes such as dust growth, radial drift, planetesimal formation, growth rates, and planetary migration.

Simplified disc models typically feature a simple power-law description of the gas surface density and disc temperature, based on stellar irradiation only \citep{weidenschilling_aerodynamics_1977, Hayashi_1981, Chiang_2010}.
However, gas accretion onto the central star gives rise to a second source of heating in the disc, accretion heating (referred in this work also as viscous heating). This type of heating is mediated by turbulence dissipation associated with momentum transport, that also drives the accretion onto the star and is typically parametrised through the parameter $\alpha_{\nu}$ \citep{shakura_black_1973}.
Simulations have shown that the temperature profile in the inner disc can be significantly elevated when stellar accretion rates are high \citep{Garaud_2007,Davis_2005,Bitsch_2013}, even up to a few tens of AU during accretion outbursts \citep{Cieza_2016}, potentially leading to significant temperature changes in planet-forming regions. Therefore, it is important to include accretion heating in the disc model.
The main uncertainties when modelling the temperature profile due to accretion heating lie in the position of the heating layer and the dust opacity. Recent studies involving non-ideal magnetohydrodynamics (MHD) show that accretion heating could take place several scale heights above the midplane \citep{Mori_2019, Bethune_2020}, contrary to the standard model of midplane accretion heating \citep{Hubeny_1990, Menou2004}.
The effective dust opacity is fundamental to determine the disc temperature profile, as the midplane temperature can change by up to a factor of 2 when considering the opacity dependency on frequency and the scattering process of dust particles \citep{Dullemond_2002, inoue_2009}.
Here we make use of a simplified grain opacity model that considers small dust grains as the dominant source of opacity, as explained in Appendix \ref{app:visc_heating}, however we note that the presence of these particles could be reduced by settling and growth processes \citep[e.g.][]{Kondo_2023}.

Given the relevance of the disc structure for planet formation, in this work, we consider two different models to describe the disc temperature profile that encompass the aforementioned processes.
One end-member model only considers heating through stellar irradiation (model \texttt{mod:irrad}). The other alternative model includes a certain degree of viscous heating in the inner part of the protoplanetary disc (model \texttt{mod:surfheat} describes a moderate degree of viscous heating and \texttt{mod:midheat} describes strong viscous heating).

For the irradiated disc, the aspect ratio is given by
\begin{equation}
    \label{eq:H_R_irr}
\left(\frac{H}{r}\right)_{\mathrm{irr}} = 0.024 \left(\frac{M_{\star}}{M_{\odot}}\right)^{-4/7} \left(\frac{L_{\star}}{L_{\odot}}\right)^{1/7} \left(\frac{r}{\mathrm{AU}}\right)^{2/7}\,,
\end{equation}
following  \citet{Garaud_2007}, \citet{ida_radial_2016}, and \citet{liu_super-earth_2019}.

To model the effect of the deposition of accretion heat in the inner disc, we explore the heating efficiency as a function of the height of the heating layer above the midplane. 
The resulting prescription for the disc aspect ratio due to accretion heating takes the form
\begin{align}
\label{eq:H_R_visc}
  \left(\frac{H}{r}\right)_{\mathrm{visc}} &=
     0.019
     \left( \frac{\epsilon_{\rm el}}{10^{-2}} \right)^{1/10}
     \left( \frac{\epsilon_{\rm heat}}{0.5} \right)^{1/10}
     \left( \frac{\alpha_{\nu}}{10^{-2}} \right)^{-1/10}
     \notag\\
     &\left(\frac{Z}{0.01} \right)^{1/10}
     \left( \frac{a_{\rm gr}}{0.1\,{\rm mm}} \right)^{-1/10}
     \left(\frac{\rho_{\rm gr}}{1\,{\rm g\,/cm}^3} \right)^{-1/10}   
     \notag\\
     & \left( \frac{\dot M_{\star}}{10^{-8} M_\odot / {\rm yr}} \right)^{1/5}
     \left( \frac{M_\star}{M_\odot}\right)^{-7/20}
     \left( \frac{r}{\rm au} \right)^{1/20}
     \,.
\end{align}
Here, $\alpha_{\nu}$ is the viscous parameter that models the viscosity of the disc, which we nominally assume $\alpha_{\nu} = 10^{-2}$, consistent with the observed $\dot{M}_{\star}$ and inferred $\Sigma_{\mathrm{gas}}$ \citep{hartmann_accretion_1998, Andrews_2009}. This value is also compatible with turbulence driven by magnetorotational instability (MRI) in ideal MHD simulations \citep{Balbus_1998}.     
The parameter $\epsilon_{\rm heat}$ expresses the conversion efficiency of accretion stress into accretion heating at a given elevation above the disc midplane, which is expressed through $\epsilon_{\rm el}$, that decreases at higher altitudes. Standard midplane heating corresponds to $\epsilon_{\rm el}=1$ \citep{Oka_2011,ida_radial_2016}, while non-ideal MHD simulations argue crudely for $\epsilon_{\rm el}=10^{-2}$ or lower \citep{Mori_2019,Mori_2021}.
A complete derivation and an expanded discussion can be found in the Appendix \ref{app:visc_heating}.
The parameters $Z$, $a_{\rm gr}$, and $\rho_{\rm gr}$ express, respectively, the dust-to-gas ratio, radius, and density of the grains that provide the opacity in the disc, and are set to their nominal values. Table \ref{tab:params} summarises the standard parameters used in the simulations.

\begin{table}[t!]
\centering
\caption{Nominal values for simulation parameters. In parentheses other tested values of the given parameter during the study.}
\resizebox{\columnwidth}{!}{
    \begin{tabular}{ cccc }
        \hline
        \hline
        Quantity & Value & Meaning & Disc model\\
        \hline
        $Z$ & $0.01$ & nominal dust-to-gas ratio  & all models \\
        $f_{\rm iceline}$ & $1$, ($0.5$) & pebble flux reduction at the iceline  & all models \\
        $\kappa_{\mathrm{env}}$ & $0.005 \: \mathrm{m^2/kg}$ & planet's envelope opacity    & all models\\
        $\alpha_{\nu}$  &  $10^{-2}$  &  viscous parameter & all models\\
        $\alpha_z$ &  $ 10^{-4}, (10^{-3})$   &  vertical stirring  & all models \\
        $\alpha_{\mathrm{frag}}$ &  $ 10^{-4}, (10^{-3})$   &  fragmentation alpha  & all models \\
        $v_{\mathrm{frag}}$ & $1$ m/s , ($10$ m/s)& fragmentation velocity  & all models \\
        $a_{\mathrm{gr}}$ & $0.1$ mm & dust grain size  & viscous models \\
        $\rho_{\mathrm{gr}}$ & $1 \: \rm g/cm^2$ & dust grain density  & viscous models \\
        $\epsilon_{\mathrm{el}}$ & $10^{-2}$ & heating elevation parameter &   \texttt{mod:surfheat} \\
        $\epsilon_{\mathrm{el}}$ &  $1$ & & \texttt{mod:midheat} \\
        $\epsilon{_\mathrm{heat}}$ & $0.5$  & heating dissipation efficiency &  \texttt{mod:surfheat} \\
        $\epsilon_{\mathrm{heat}}$ &  $1$ & & \texttt{mod:midheat} \\
        \hline
    \end{tabular}
}
\label{tab:params}
\end{table}

For the \texttt{mod:midheat} and \texttt{mod:surfheat} discs, the disc aspect ratio at any time step is described both by viscous heating and irradiation, depending on which process dominates at a given location.
At each position and for each time step, the disc aspect ratio is computed as the maximum between the two values: 
\begin{equation}
\label{eq:H_R_visc_final}
    \frac{H}{r} = 
    \max \left[\left( \frac{H}{r} \right)_{\mathrm{visc}}, \left( \frac{H}{r} \right)_{\mathrm{irr}} \right].
\end{equation}

\begin{figure}[t!]
    \centering
    \includegraphics[width=\linewidth]{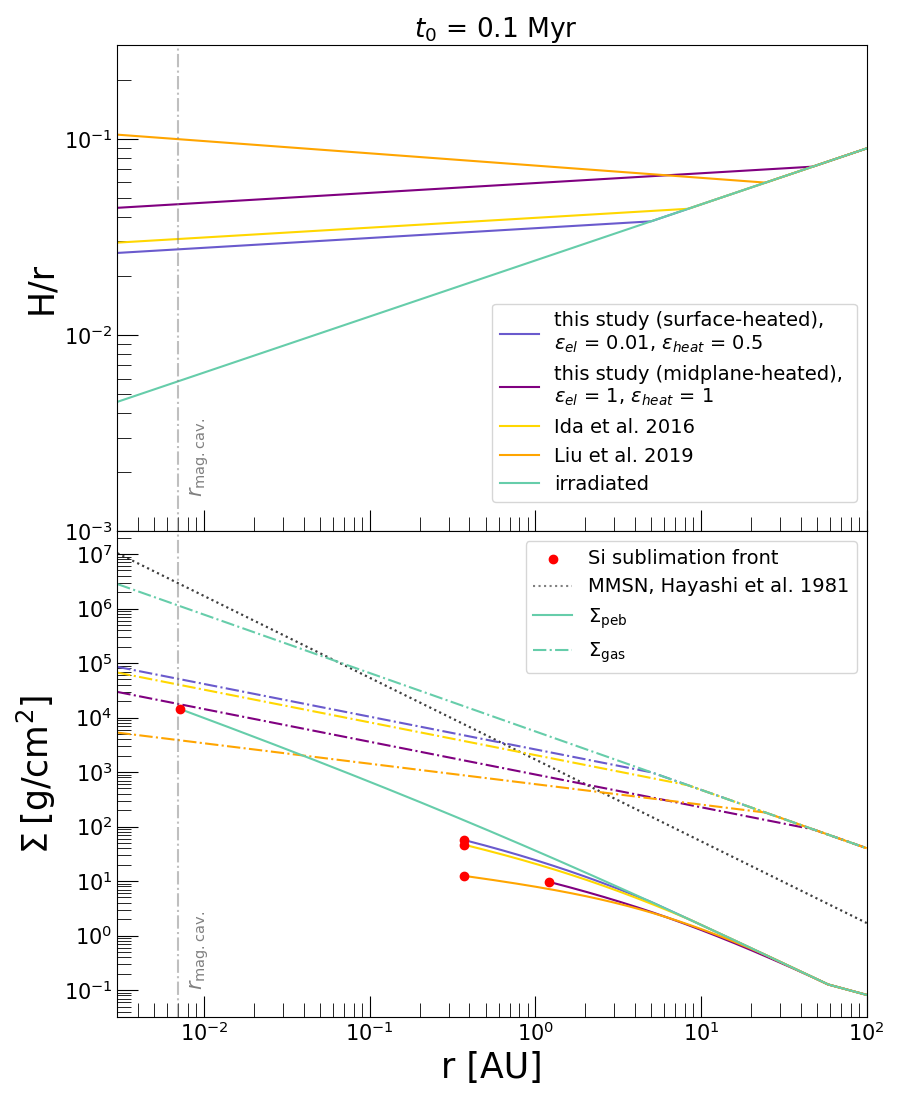}
    \caption{
    Disc gas scale height aspect ratio (upper panel) and  surface density of gas and dust (lower panel, solid and dashed-dotted lines respectively) for different disc models, at initial time $t_0 = 0.1$\,Myr (Section\,\ref{subsect:gas_disc}). 
    The green line shows a purely irradiated disc. 
    Purple and blue curves show discs, respectively, with midplane or surface accretion heating.
    We plot, for comparison, in yellow and orange the viscous disc models in \citet{ida_radial_2016} and \citet{liu_super-earth_2019}.
    We mark with a grey dashed-dotted vertical line the position of the magnetospheric cavity 
    at the initial time (Eq.\,\ref{eq:r_mag_cav}).
    In the lower panel, the grey dotted line shows the gas surface density of the minimum mass solar nebula for reference \citep{Hayashi_1981}, while red dots mark the silicate sublimation line (Eq.\,\ref{eq:r_ice_visc_fin}).
    }
    \label{fig:disc_quantities}

\end{figure}

To model the gas component of the protoplanetary disc, we use a simple $\alpha$-viscosity prescription \citep{shakura_black_1973}, where the viscosity is given by $\nu = \alpha_{\nu} c_{\mathrm{s}} H$. 
Here, $H$ is the gas scale height and $c_{\mathrm{s}}= H \Omega_{\mathrm{K}}$ the sound speed.
In a steady-state disc \citep{Pringle_1981}, the viscosity ($\nu$), gas accretion rate onto the star ($\Dot{M}_{\star}$), and gas surface density ($\Sigma_{\mathrm{gas}}$) are related through
\begin{equation}
    \label{eq:M_dot}
    \Sigma_{\mathrm{gas}} = \frac{\Dot{M}_{\star}}{3 \pi \alpha_{\nu} H^2 \Omega_{\mathrm{K}}},
\end{equation}
allowing us to derive the gas surface density at each time step by combining Eq.\,(\ref{eq:M_dot}) with Eq.\,(\ref{eq:H_R_irr}) for the irradiated disc model and Eq.\,(\ref{eq:H_R_visc_final}) for the viscous disc models, as shown in Fig.\,\ref{fig:disc_quantities}.
We model the gas accretion onto the central star as a function of time using the upper limit fit in \citet{hartmann_accretion_2016}:
\begin{equation}
    \label{eq:sigma-gas_dot}
    \log_{10}{\left(\frac{\Dot{M}_{\star}}{M_{\odot}/\mathrm{yr}}\right)} = -1.32 -1.07 \log_{10}{\left(\frac{t}{\mathrm{yr}}\right)} \: .
\end{equation}
This is a pragmatic choice, in order to have a decrease in the stellar accretion rate roughly consistent with observations \citep{bitsch_structure_2015,liu_super-earth_2019}.

Finally, the inner edge of the disc is set at the magnetospheric cavity radius as in \citet{frank_accretion_2002}, \citet{armitage_astrophysics_2010}, \citet{liu_dynamical_2017}
\begin{multline}
    \label{eq:r_mag_cav}
    r_{\mathrm{mag, cav}} = \left(\frac{B_{\star}^4 R_{\star}^{12}}{4GM_{\star}\Dot{M}_{\star}^2}\right)^{1/7}  \\ 
     \simeq 0.0167 \: \left(\frac{B_{\star}}{1 \: \rm kG}\right)^{4/7} \left(\frac{R_{\star}}{R_{\odot}}\right)^{12/7} \left(\frac{M_{\star}}{M_{\odot}}\right)^{-1/7} \left(\frac{\Dot{M}_{\star}}{10^{-8} M_{\odot}/ \rm yr}\right)^{-2/7} \rm AU, 
\end{multline}
where $B_{\star}$ is the stellar magnetic field, $R_{\star}$ the stellar radius, $M_{\star}$ the stellar mass and $\Dot{M}_{\star}$ the accretion rate on the central star (Eq.\,\ref{eq:M_dot}).
 We set our nominal values to $B_{\star} = 1 \: \rm kG$, representative of a typical solar mass T Tauri star \citep[e.g.][]{Johns-Krull_2007}, $R_{\star} = 1 \: R_{\odot}$, and $M_{\star}= 1 \: M_{\odot}$.

\subsection{Pebble disc model}
\label{sec:pebble_disc}
The following section is devoted to the description of pebbles in the disc, namely how we determine pebbles sizes, their drift velocity, the resulting pebble flux that is available for the planet to accrete and how we treat the presence of the water iceline.

\subsubsection{Pebble size}
\label{subsect:pebble_size}
In our work, we use a single population of pebbles whose size is either drift or fragmentation limited. The drift-limited size is the maximum dimension that the dust can grow to before starting to significantly drift inwards, while the fragmentation limit is the maximum size that can be reached before the collisions between dust grains become disruptive. Typically, the fragmentation barrier is lower than the drift barrier, although this could no longer be true in the outer disc and for sufficiently high fragmentation velocities. 

The fragmentation limited Stokes number is given by \citep{ormel_closed-form_2007}:
\begin{multline}
    \label{eq:St_frag}
        \mathrm{St_{frag}} = \frac{v_{\mathrm{frag}}^2}{3 \alpha_{\mathrm{frag}} c_{\mathrm{s}}^2} \\
        = 0.015 \left(\frac{\alpha_{\mathrm{frag}}}{10^{-3}}\right)^{-1} \left(\frac{v_{\mathrm{frag}}}{10 \:\mathrm{m/s}}\right)^{2} \left(\frac{H/r}{0.05}\right)^{-2} \left(\frac{r}{\mathrm{AU}}\right)\left(\frac{M_{\star}}{\mathrm{M_{\odot}}}\right)^{-1},
\end{multline}
with $v_{\mathrm{frag}}$ dust fragmentation velocity, $\alpha_{\mathrm{frag}}$ dust turbulence velocity coefficient, and $c_{\mathrm{s}}$ isothermal speed of sound.
In what follows, we will set $\alpha_{\rm frag} = \alpha_{\rm z} = 10^{-4}$ representing a low-turbulence midplane, as suggested in \citet{jiang_grain-size_2024} based on protoplanetary disc observations assuming a $1 \: \rm m/s$ fragmentation velocity.

The drift limited Stokes number is derived by equating the drift timescale (Eq.\,\ref{eq:t_d}) with the growth timescale (Eq.\,\ref{eq:t_g}). The former one is simply given by
\begin{equation}
    \label{eq:t_d}
    t_{\mathrm{drift}} = \frac{r}{v_r},
\end{equation}
with $v_r$ pebble drift velocity \citep[e.g.][]{nakagawa_settling_1986, guillot_filtering_2014,ida_radial_2016}
\begin{equation}
    \label{eq:v_r_maintext}
    v_r = - 2 \frac{\mathrm{St}}{1+\mathrm{St^2}} \eta v_{\mathrm{K}} + \frac{\mathrm{1}}{1+\mathrm{St^2}} \eta v_{\mathrm{K}} u_{\nu},
\end{equation}
where $u_{\nu} \approx -\nu/r \approx - \alpha_{\nu} (H/r)^2 v_{\mathrm{K}}$ is the radial viscous diffusion velocity. The first term in Eq.\,(\ref{eq:v_r_maintext}) is typically the dominant one, although, in the inner disc, the diffusion velocity can become relevant.
For simplicity, in the calculation of the drift limited Stokes numbers, we neglect the second term in Eq.\,(\ref{eq:v_r_maintext}) and approximate $\mathrm{St^2}+1 \approx 1$, giving as radial drift velocity $v_r \approx -2\mathrm{St} \eta v_{\mathrm{K}}$.

The growth timescale is defined as the ratio between the grain dimension and the grain growth rate
\begin{equation}
\label{eq:t_g}
    t_{\mathrm{growth}} = \frac{a_{\rm gr}}{\Dot{a}_{\rm gr}} = a_{\rm gr} \frac{4}{\sqrt{3}} \frac{\rho_{\mathrm{gr}}}{\rho_{\mathrm{gas}}} \frac{1}{\mathrm{St} \: c_{\mathrm{s}} \: Z}, 
\end{equation}
where $\rho_{\mathrm{gr}}$ is the grain density and $\rho_{\mathrm{gas}}$ the gas volume density. We used here the prescription of \citet{birnstiel_simple_2012} to determine the grain growth rate (see Appendix \ref{app:st_drift} for a more detailed calculation).
To express the dimension of the grain $a_{\rm gr}$ in terms of Stokes number, we need to know the drag regime that the grains are subjected to: Epstein drag (if $a_{\rm gr} < 9/4 \: \lambda_{\mathrm{mfp}}$) or Stokes drag (if $a_{\rm gr} > 9/4 \: \lambda_{\mathrm{mfp}}$). In the first case the Stokes number depends linearly on the grain dimension, in the latter the dependence is quadratic.
The grain dimension $a_{\rm gr}$ can be related to the Stokes number in each drag regime as \citep{ida_radial_2016}
\begin{equation}
\label{eq:st_ep_reg}
\begin{cases}
   \mathrm{St_{Ep}} = \frac{\rho_{\mathrm{gr}} \Omega_{\mathrm{K}}}{\rho_{\mathrm{gas}} c_{\mathrm{s}}} a_{\rm gr} & a_{\rm gr}<\frac{9}{4} \lambda_{\mathrm{mfp}}, \\
   \mathrm{St_{St}} = \frac{4}{9}\frac{\rho_{\mathrm{gr}}}{\rho_{\mathrm{gas}}}\frac{\Omega_{\mathrm{K}}}{c_{\mathrm{s}} \lambda_{\mathrm{mfp}}} a_{\rm gr}^2 & a_{\rm gr}>\frac{9}{4} \lambda_{\mathrm{mfp}}, 
\end{cases}
\end{equation}
where $\lambda_{\mathrm{mfp}} = (\mu m_{\mathrm{p}})/(\sigma_{\mathrm{H}} \rho_{\mathrm{gas}})$ is the gas mean free path.
By equating the growth and the drift timescales in the two respective drag regimes, we obtain an explicit expression for the Stokes number of the particle (see Appendix \ref{app:st_drift} for the detailed calculation)

\begin{equation}
    \mathrm{St_{drift, Ep}} = \sqrt{\frac{\sqrt{3} \epsilon_{\mathrm{p}}F_{\mathrm{peb}}}{32 \pi \Sigma_{\mathrm{gas}}  \eta^2 v_{\mathrm{K}}r}},
\end{equation}
\begin{equation}
    \label{eq:st_drift_st}
    \mathrm{St_{drift, St}} = \left(\frac{48\pi}{\sqrt{3}} \frac{\Sigma_{\mathrm{gas}}}{F_{\mathrm{peb}}} \sqrt{\frac{\rho_{\rm gr} \lambda_{\mathrm{mfp}}}{\rho_{\mathrm{gas}}H}} \frac{ \eta^2 v_{\mathrm{K}}^2}{\epsilon_{\mathrm{p}}\Omega_{\mathrm{K}}}\right)^{-2/3},
\end{equation}
where $\epsilon_{\mathrm{p}} = 0.5$ is a parameter that we introduced to account for the coagulation efficiency between grains and $F_{\mathrm{peb}}$ is the pebble flux.

\begin{figure*}
    \centering
    \includegraphics[width = \textwidth]{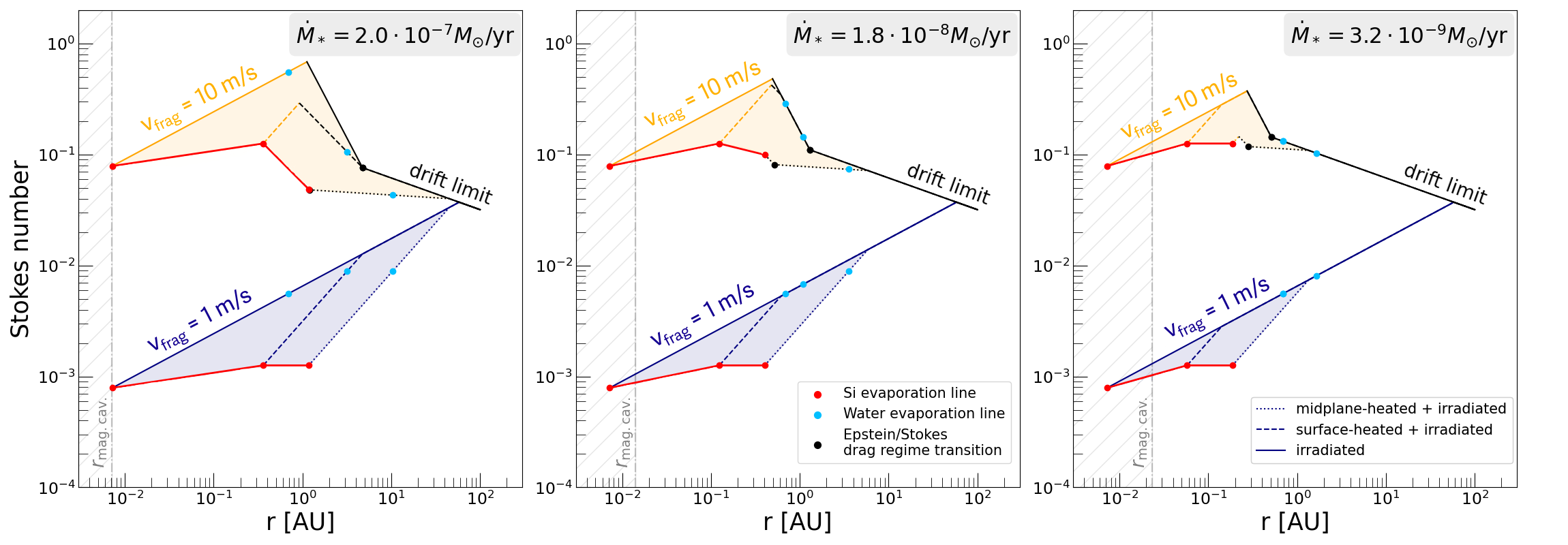}
    \caption{
    Stokes number of pebbles as a function of position, for different gas accretion rates in the different panels, corresponding to $t \simeq 0.1$ Myr (left), $t \simeq 1.0$ Myr (central), and $t \simeq 5.0$ Myr (right).
    The blue branch shows the Stokes number for $v_{\mathrm{frag}} = 1 \: \rm m/s$, while the orange branch for $v_{\mathrm{frag}} = 10 \: \rm m/s$. The  black line marks where the Stokes number is limited by drift rather than fragmentation.
    The different line styles represent the three different disc models that we used: purely irradiated disc (\texttt{mod:irrad}, solid line), 
    moderately viscously heated inner disc (\texttt{mod:surfheat}, dashed line) and 
    efficiently viscously heated inner disc (\texttt{mod:midheat}, dotted line). 
    In each panel, the light blue dots mark the water evaporation front, while the red dots, with corresponding connecting lines, mark the location of the Si evaporation front.
    The black dots represent the location at which particles enter the Stokes drag regime. 
    Finally, the grey dashed-dotted line represents the location of the inner magnetospheric cavity given by Eq.\,(\ref{eq:r_mag_cav}), that we consider as inner disc edge. 
    Low fragmentation velocities near $1 \: \rm m/s$, and increased levels of viscous heating in young discs, reduce Stokes numbers in the inner disc.}
    \label{fig:st_frag}
\end{figure*}
To determine the final size of the pebble at each time step and position in the disc, we first determine if a pebble is in the Stokes or Epstein drag regime. 
We then pick the minimum size between the drift-limited and the fragmentation-limited sizes:
\begin{equation}
    \label{eq:st_final}
    \mathrm{St} = \min [\mathrm{St_{drift}}, \mathrm{St_{frag}}]
\end{equation}
The Stokes numbers of the pebbles in our disc models as a function of positions are shown in Fig.\,\ref{fig:st_frag}, where the three panels represent three different gas accretion rates corresponding to $t \simeq 0.1, 1, 5$ Myr respectively. 
The two different colours identify the fragmentation limited sizes for $v_{\mathrm{frag}} = 1\: \rm m/s$ (blue) and $v_{\mathrm{frag}} = 10 \: \rm m/s$ (orange), while the black lines identify drift limited sizes. 
Starting in the outer disc, pebble sizes are initially drift limited, but as dust particles drift inwards their size gets set by the fragmentation limit. 
This transition occurs at wider orbital radii
in case of low $1 \: \rm m/s$ fragmentation velocities (blue branch) compared to higher fragmentation velocities (orange branch), leading to smaller maximal pebbles Stokes numbers for smaller $v_{\mathrm{frag}}$. Fig.\,\ref{fig:R_peb_r} in Appendix \ref{app:st_drift} shows the corresponding pebble size in cm, derived through Eq.\,(\ref{eq:st_ep_reg_app}) assuming compact spheres. In this work we do not take dust porosity into account.

Focusing on the $1 \: \rm m/s$ fragmentation branch, the solid, dashed and dotted lines represent, respectively,
a purely irradiated disc model (\texttt{mod:irrad}), 
a disc dominated by surface heating (\texttt{mod:surfheat}), and a disc dominated by midplane heating (\texttt{mod:midheat}). 
Strong viscous heating (dotted lines) leads to an increased midplane temperature that heavily limits the size of fragmentation-limited pebbles (cfr. Eq.\,\ref{eq:St_frag}). 
This effect decreases as the disc evolves with time (left to right panel). Below accretion rates of approximately $\dot M_{\star} \approx 5 \cdot 10^{-9}$\,M$_\odot$/yr, the region outside $1$\,AU is firmly dominated by irradiation (solid line).  
In the high fragmentation velocities case ($10 \: \rm m/s$, orange branch), particles are larger and drift-limited even up to 1 AU and can enter the Stokes drag regime (black dots) in the terrestrial planet formation region. In this regime, the Stokes number remains below unity, but, as seen in Fig.\,\ref{fig:R_peb_r}, the sizes increase up to small planetesimals \citep{Okuzumi_2012}.

The red dots and lines mark the Si sublimation front for each disc model, where we expect most rock-forming material to sublimate. 
In the cold irradiated disc, the Si sublimation front can be close to, or even within, the magnetospheric cavity radius (grey dashed region, Eq.\,\ref{eq:r_mag_cav}). 
Modelling the very inner disc regions ($T>1000$\,K), where thermal ionisation by alkali elements may be sufficient to drive MRI-turbulence in the very inner disc \citep{Desch_2015}, is outside the scope of this study.

To summarise, within approximately the water sublimation line (blue dots in Fig.\,\ref{fig:st_frag}), pebble sizes become strongly dependent on the assumed fragmentation velocity and the disc midplane temperature, potentially leading to orders of magnitude difference in the Stokes number of particles in the inner disc.
Going forward, we will assume a nominal fragmentation velocity of $1 \: \rm m/s$ (cfr. Table \ref{tab:params} for nominal simulation parameters), which is the conventional fragmentation velocity assumed for rocky particles \citep{guttler_outcome_2010}. Nevertheless, since this is a critical model parameter, we will also investigate higher fragmentation velocities as advocated by some groups \citep{yamamoto_examination_2014,Kimura_2015}. 

\subsubsection{Pebble scale height}
\label{subsect:pebble_scaleheight}

To model the pebble reservoir in the disc, we consider the pebbles to be distributed in a disc with scale height given by \citep{youdin_particle_2007}
\begin{equation}
    \label{eq:H_peb}
    H_{\mathrm{peb}} = \sqrt{\frac{\alpha_z}{\alpha_z + \mathrm{St}}} H,
\end{equation}
where $\alpha_z$ is the turbulent stirring that lifts the particle up from the midplane, $\mathrm{St}$ is the particle Stokes number (see Section \ref{subsect:pebble_size}), and $H$ is the gas scale height. 
We set as our nominal value $\alpha_{\rm z}  = 10^{-4}$ representing a low-turbulence midplane,
as supported by recent observations of dust scale heights \citep{pinte_2016}, estimates of turbulent line broadening \citep{flaherty_2020}, and fragmentation-limited particle sizes \citep{jiang_chemical_2023}.

\subsubsection{Pebble flux}
\label{subsect:pebble_flux}
The pebble mass reservoir in the disc strongly regulates the outcome of planet formation. Discs with a larger dust-to-gas ratio (metallicity), or radial extent, harbour a larger or more sustained flux of radially inwards-drifting pebbles \citep{lambrechts_forming_2014}. 
Even small differences within a factor two in the pebble flux can result in inner systems composed of terrestrial planets or super-Earths \citep{lambrechts_formation_2019}.
In this work we aim to focus on the subset of discs that effectively generate a substantial pebble flux capable of forming gas giant planets in Jupiter-like orbits.
Such discs may be relatively rare \citep{van_der_marel_stellar_2021}, but are of prime interest in understanding our Solar System architecture and the origin of cold giant systems with super-Earths.
To do so, we have chosen a simplified model in which the pebble flux is set in the outer disc, where we assume a stream of small dust grains carried along with the gas flow, with a solar dust-to-gas ratio ($Z=0.01$). As these particles drift inwards, they then undergo coagulation and fragmentation up to the limit sizes discussed in Section \ref{subsect:pebble_size}.
Alternative approaches can be found in the literature that are analytical \citep{lambrechts_forming_2014,johansen_how_2019,gurrutxaga_formation_2024}, simplified numerical \citep[e.g.][]{appelgren_disc_2023}, or full hybrid/full coagulation simulations \citep{birnstiel_testing_2010}. 

To account for the total mass reservoir that our planetary embryos can accrete, we define a nominal flux $F_0$, which is constant in space and decreases with time as a fraction of the gas accretion rate onto the central star 
\begin{equation}
    \label{eq:F_0}
    F_0 = Z \: \Dot{M}_{\star}f_{\rm iceline},
\end{equation}
with $Z = 0.01$ initial dust-to-gas ratio.
We also include the possibility of a flux reduction when crossing the iceline expressed by a multiplicative factor $f_{\rm iceline}$, which is set to 1 (no reduction) for our nominal simulations.
The nominal flux $F_0$ is calculated at each time step by getting the corresponding gas accretion rate, $\Dot{M}_{\star}$, through Eq.\,\ref{eq:sigma-gas_dot} assuming a constant dust-to-gas ratio $Z$ of small grains in the outer disc.

Since we want to investigate what the role is of the simultaneous growth of outer giants and inner planets, we simulate systems consisting of n planets, where the i-th planet in the system accretes pebbles from a reduced flux

\begin{equation}
\label{eq:F_p}
F_{\mathrm{peb,i}} = \Pi_{k=0}^{i-1} F_0(1 - f_k),
\end{equation}\\
where $f_k$ is the fraction of pebbles accreted by each of the planets outside the i-th planet's orbit, and is given by

\begin{equation}
\label{eq:ff}
f_k = \frac{\Dot{M}_{\mathrm{k}}}{F_{\mathrm{peb, k}}},
\end{equation}\\
with $\Dot{M}_{\mathrm{k}}$ mass accretion rate on the k-th planet, as expressed by the pebble accretion prescriptions in Section \ref{subsect:pebble_accretion} and $F_{\mathrm{peb, k}}$ flux on the k-th planet.
The idea is that the pebble reservoir coming from the outer disc and drifting inwards gets reduced by the chain of accreting outer embryos before reaching the innermost embryos.

\subsubsection{Sublimation fronts}
\label{subsect:iceline}
We determine the location of the sublimation fronts in the disc midplane by calculating the position at which the midplane temperature of the disc is equal to the corresponding sublimation temperature of a given molecule.
For the purpose of our study, the relevant sublimation fronts to be considered are the water and silicates sublimation fronts, respectively found at $T_{\mathrm{H_2O}} = 170 $ K and $T_{\mathrm{Si}} = 1200 $ K \citep{Hayashi_1981}.
Using the relation
\begin{equation}
\label{eq:iceline_calc}
    \frac{H}{r} = \frac{c_{\mathrm{s}}}{v_{\mathrm{K}}},
\end{equation}
we can explicitly derive the positions of the water and Si sublimation fronts (cfr. Section \ref{subsect:role_iceline}), which for the irradiated model are time-independent, thus situated at a fixed location, while for the viscously heated models are moving inwards with time as the inner disc cools down.

 Similarly to what happens for the disc aspect ratio, the position of the iceline in the irradiated case is exclusively given by Eq.\,\ref{eq:iceline_irr}, while in the viscous cases is set by
\begin{equation}
\label{eq:r_ice_visc_fin}
    r_{\mathrm{H_2O/Si}} = \max [r_{\mathrm{H_2O/Si, visc}}, r_{\mathrm{H_2O/Si, irr}}]
\end{equation}
To model the volatile mass loss of the inwards drifting pebbles, we include the option of reducing the pebble flux by a factor 2 when crossing the water sublimation front, based on a solar ice-to-dust mixture \citep{Lodders_2003}. We also include the possibility of an increase in vertical turbulence when crossing the iceline \citep{jiang_grain-size_2024}, as well as a reduction in the pebble size. In Section. \ref{subsect:role_iceline} we analyse the impact of the possible reduction in flux and particle size and increase in vertical stirring on the outcome of the simulations.

\subsection{Planetary growth}
\label{subsec:plangrowth}

\subsubsection{Planetary embryos}
\label{subsect:m0_embryos}
The starting planetary embryos that will accrete pebbles are assumed to have been formed in the early stages (first $10^5$ yrs) of the disc through streaming instability. We thus model our initial mass function for planetary embryos as in \citet{Liu_2020}

\begin{equation}
\label{eq:M0_pla}
  M_{0,\mathrm{pla}} = 2\cdot 10^{-4}
                 \frac{f}{400}
                 \left(\frac{H/r}{0.04} \right)^{3/2}
                 \left( \frac{\Sigma_{\rm gas}}{1700\,{\rm g/cm}^2} \right)^{3/2}
                 \left( \frac{r}{\rm AU} \right)^3
                 \, M_{\oplus} \, , 
\end{equation}\\ 
where $f$ is a multiplicative factor that expresses how much the most massive planetesimals exceed the characteristic planetesimal mass of the measured size distribution. Here we used $f = 400$ as in \citet{Liu_2020}.

\subsubsection{Pebble accretion}
\label{subsect:pebble_accretion}
Pebble accretion can occur in different regimes, namely 3D or 2D and Bondi or Hill.
Whether an embryo accretes in the 3D or the 2D pebble accretion regime depends on how the pebble accretion radius compares to the pebble scale height in the disc. If the accretion radius is smaller than the pebble scale height, the planetary embryo accretes spherically from a sphere of radius $R_{\mathrm{acc}}$ of density $\rho_{\mathrm{peb}}$: this is the case for the 3D accretion. If the accretion radius is larger than the pebble scale height, the embryo accretes from a circular section of radius $R_{\mathrm{acc}}$ and surface density  $\Sigma_{\rm peb}$: this is the case for the 2D accretion.
The Bondi/Hill regime transition, instead, is marked by the dominant term in the pebble approach velocity to the embryo. If the pebble velocity is mainly set by the ``shear'' term due to the star's gravity, the accretion happens in the Hill regime, if the velocity is dominated by the ``headwind'' term due to the gas drag, the accretion happens in the Bondi regime.

The 3D and 2D regime accretion equations, regardless of whether they are in the Bondi or Hill regime, are given by
\begin{align}
    \label{eqz:peb_acc_3D_maintext}
    \Dot{M}_{\mathrm{3D}} &=  \pi R_{\mathrm{acc}}^2 \rho_{\mathrm{peb}} v_{\rm acc} = \pi R_{\mathrm{acc}}^2 \frac{\Sigma_{\mathrm{peb}}}{\sqrt{2 \pi}H_{\mathrm{peb}}}  v_{\mathrm{acc}}, \\
    \label{eqz:peb_acc_2D_maintext}
    \Dot{M}_{\mathrm{2D}} &= 2 R_{\mathrm{acc}} \Sigma_{\mathrm{peb}}  v_{\mathrm{acc}}, 
\end{align}
where $R_{\rm acc}$ is the pebble accretion radius and $v_{\mathrm{acc}}$ denotes the velocity with which the pebble approaches the embryo \citep{ormel_effect_2010,lambrechts_rapid_2012}. Here, we made use of the midplane approximation to convert the pebble volume density into a pebble surface density.
The product between accretion radius and velocity ($R_{\rm acc} ^2 v_{\rm acc}$) can be retrieved by equating the settling timescale with the encounter timescale 
\citep{lambrechts_formation_2019}.
Since in the 3D regime the dependency of the accretion rate is directly set by the product $R_{\rm acc} ^2 v_{\rm acc}$, the final accretion prescription takes the same form both in the Bondi and the Hill regime. In contrast, in the 2D case, the accretion rate depends on the product $R_{\rm acc} v_{\rm acc}$, therefore an explicit expression for both accretion radius and velocity is needed, leading to two different equations for the Bondi and the Hill regimes \citep[see][]{Ormel_2017}.
Substituting the values of accretion radius and velocity into Eqs.\,(\ref{eqz:peb_acc_3D_maintext}, \ref{eqz:peb_acc_2D_maintext}), we find the following expression for accretion in the 3D and 2D Bondi and Hill regimes
\begin{align}
    \label{eq:3D_acc}
    & \Dot{M}_{\mathrm{3D}} = 6 \pi R^3_{\mathrm{H}}\mathrm{St} \Omega_{\mathrm{K}}\rho_{\mathrm{peb}},
     \\
     \label{eq:2D_acc_B}
    & \Dot{M}_{\mathrm{2D, Bondi}} = 2 \sqrt{2 G M \mathrm{St} \frac{ v_{\mathrm{HW}}}{\Omega_{\mathrm{K}}}} \Sigma_{\mathrm{peb}},
         \\
     \label{eq:2D_acc_H}
    & \Dot{M}_{\mathrm{2D, Hill}} = 3(4\mathrm{St})^{2/3}R^{2}_{\mathrm{H}}\Omega_{\mathrm{K}} \Sigma_{\mathrm{peb}},
\end{align}
with $R_{\mathrm{H}} = \left(\frac{M}{3M_{\star}}\right)^{1/3}r$ Hill radius, $\Omega_{\mathrm{K}}$ Keplerian frequency, and $v_{\rm HW} \simeq \eta v_{\rm K}$ headwind velocity.
The transition between the 3D and 2D Bondi or Hill regimes happens when the accretion radius is larger than the pebble scale height
\begin{equation}
    \label{eqz:2d3d_trans}
    R_{\mathrm{acc}}  \geq \frac{2\sqrt{2\pi}}{\pi} H_{\mathrm{peb}}.
\end{equation}
This can be translated in an embryo mass criterion, that yields
\begin{align}
    \label{eqz:M_3D_2D_trans_reg}
    & M_{\mathrm{3D \rightarrow 2D, Hill}} = 6
    \left(\frac{\sqrt{2 \pi}}{\pi} \right)^3 \left(\frac{H}{r}\right)^3
    M_{\star}
    \alpha_z^{3/2} \mathrm{St}^{-5/2}, \\
    \label{eqz:2DB_trans}
    & M_{\mathrm{3D \rightarrow 2D, \:Bondi}} = \frac{2}{\pi} \left(\frac{d \ln P}{d \ln r}\right) \left(\frac{H}{r}\right)^4 M_{\star}
     \alpha_z {\rm St^{-2}}.
\end{align}
We notice that the vertical stirring of the pebbles in the disc ($\alpha_z$) is a fundamental parameter for the accretion efficiency because it is encoded in the transition condition between 3D and 2D accretion regimes, thus determining whether the embryos accrete in the (generally) more efficient 2D regime or in the (generally) less efficient 3D regime.

\subsubsection{Pebble isolation mass}
Planetary embryos grow through pebble accretion until they are massive enough to gravitationally perturb the gas in the disc, creating a pressure bump that halts the inwards drift and accretion of pebbles. This limit mass is the so-called pebble isolation mass, which also marks the beginning of the planet's gas accretion phase.
We use the prescription in \citet{johansen_forming_2017} to model the pebble isolation mass
\begin{equation}
    \label{eq:peb_iso}
    M_{\mathrm{iso}} \simeq \left(\frac{H}{r}\right)^3 \frac{\partial \ln P}{\partial \ln R} M_{\star} \approx 20 \left(\frac{H/r}{0.05}\right)^3 \left(\frac{M_{\star}}{M_{\odot}}\right) M_{\oplus}.
\end{equation}
In this model the pebble isolation mass is entirely determined by the aspect ratio and the pressure profile of the disc, however some studies report that the pebble isolation mass could increase in turbulent discs \citep{Bitsch_2018, Ataiee_2018}. 

In our nominal model, once the planet has reached isolation mass, we assume that the pressure bump blocks all the material outside the planet's orbit, preventing any other solid from drifting past it and accrete on the innermost planets.
However, some studies \citep[e.g.][]{Weber_2018, Haugbolle_2019, stammler_leaky_2023} have shown that some dust could leak through the planetary gap, possibly accreting onto the inner planets. We take this possibility into account in Section \ref{sec:leaking_dust}, where we show how different degrees of dust leaking affect the results of our simulation.

\subsection{Type I migration}
\label{subsect:typeI_mig}
The interaction between a forming planet and the protoplanetary disc leads to a change in the orbital elements of the planet as it grows (see \citet{kley_planet-disk_2012} and \citet{baruteau_planet-disk_2014} for reviews and \citet{kretke_importance_2012} for an overview on migration rates in power-law discs). Typically, as the planet is still small and embedded in the disc, it feels an asymmetric torque due to Lindblad resonances and co-rotation that leads to inward migration \citep{goldreich_disk-satellite_1980, ward_protoplanet_1997}. We model this type I migration as 
\begin{equation}
\label{eq:typeI_mig}
    \left(\frac{\mathrm{d}r}{\mathrm{d}t}\right)_{\mathrm{I}} = -c \frac{M}{M_{\star}} \frac{\Sigma_{\mathrm{gas}} r^2}{M_{\star}} \left(\frac{H}{r}\right)^{-2} v_{\mathrm{K}},
\end{equation}
where $c$ is a parameter that depends on the radial pressure and temperature profile of the disc. Following \citet{paardekooper_torque_2010} we adopt $c = 2.8$ for the isothermal regime, but other prescriptions would have a weak impact on the migration rates.
Once the planet grows large enough not to be embedded in the disc anymore, type II migration takes over \citep{ward_protoplanet_1997}. This happens approximately at the range of masses in which the planet switches to gas accretion, as we describe in Section \ref{sec:type2}.

For the purposes of this study, we stop the inwards migration of the planet (regardless of it being type I or II) at the magnetospheric cavity radius, independently of the planetary mass
\citep{Tanaka_2002, Masset_2006}.

\subsection{Type II migration}
\label{sec:type2}

As the planet grows more and more massive, it perturbs the gas in the disc creating a gap and reducing its migration rate, leading it to migrate at the same speed of the viscous accretion \citep{lin_planetary_1986}. This is commonly known as type II migration. Here we chose to adopt the physical model proposed by \citet{kanagawa_radial_2018} which shows that the torque exerted on the protoplanet is well described by the classical type I torque (that gives rise to Eq.\,\ref{eq:typeI_mig}) multiplied by the relative gap height $\Sigma_{\mathrm{gap}}/\Sigma_{\mathrm{gas}}$ carved by the planet in the disc, which can be expressed as
\begin{equation}
    \label{eq:sigma_gap_gas}
    \frac{\Sigma_{\mathrm{gap}}}{\Sigma_{\mathrm{gas}}} = \frac{1}{1+\left(\frac{M}{M_{\mathrm{gap}}}\right)^2},
\end{equation}
with $\Sigma_{\mathrm{gap}}$ gas surface density in the gap, $\Sigma_{\mathrm{gas}}$ unperturbed gas surface density, and $M_{\mathrm{gap}}$ gap transition mass, i.e. the mass required to produce a relative gap height of $0.5$. Since a relative gap height of $0.85$ is already sufficient to reach pebble isolation mass \citep{johansen_how_2019}, the gap transition mass can be expressed in terms of the pebble isolation mass as $M_{\mathrm{gap}} \approx 2.3 \: M_{\mathrm{iso}}$, leading to the following migration prescription
\begin{equation}
    \label{eq:typeII_mig}
     \left(\frac{\mathrm{d}r}{\mathrm{d}t}\right) =  \left(\frac{\mathrm{d}r}{\mathrm{d}t}\right)_{\mathrm{I}} \cdot \frac{\Sigma_{\mathrm{gap}}}{\Sigma_{\mathrm{gas}}} = \left(\frac{\mathrm{d}r}{\mathrm{d}t}\right)_{\mathrm{I}} \left(1+\frac{M}{\left(2.3 \: M_{\mathrm{iso}}\right)^2}\right)^{-1},
\end{equation}
where $M_{\mathrm{iso}}$ is the pebble isolation mass (Eq.\,\ref{eq:peb_iso}) and $\left(\frac{\mathrm{d}r}{\mathrm{d}t}\right)_{\mathrm{I}}$ the classical type I migration given by Eq.\,(\ref{eq:typeI_mig}).
Notice how this expression gives a general migration prescription smoothly connecting the type I and II migration regimes.

\subsection{Gas accretion}
\label{subsect:gas_accretion}
After the planet has reached pebble isolation mass, it starts accreting gas. The first phase is characterised by a slow Kelvin-Helmholtz envelope contraction, which we model as in \citet{Ikoma_formation_2000}
\begin{equation}
    \label{eq:env_contr}
    \Dot{M}_{\mathrm{KH}} = 10^{-5} \left(\frac{M}{10 \: M_{\oplus}}\right)^4 \left(\frac{\kappa_{\mathrm{env}}}{0.1 \: \mathrm{m^2/kg}}\right)^{-1} M_{\oplus}/\mathrm{yr},
\end{equation}
where $\kappa_{\mathrm{env}}$ is the envelope opacity, whose nominal value we set at $\kappa_{\mathrm{env}} = 0.005 \: \mathrm{m^2/kg}$. 
As the mass of the planet increases, the accretion rate in Eq.\,(\ref{eq:env_contr}) becomes higher than what the protoplanetary disc can supply in terms of gas mass. At this evolutionary stage, \citet{Tanigawa_final_2016} showed that the accretion rate of the gas onto the planet can be expressed as
\begin{equation}
    \label{eq:gas_acc_Hill}
    \Dot{M}_{\mathrm{disc}} = 0.29 \left(\frac{H}{r}\right)^{-2} \left(\frac{M}{M_{\star}}\right)^{4/3} \Sigma_{\mathrm{gas}} r^2 \Omega \frac{\Sigma_{\mathrm{gap}}}{\Sigma_{\mathrm{gas}}},
\end{equation}
where $\Sigma_{\mathrm{gap}}/\Sigma_{\mathrm{gas}}$ is the relative gap height given by Eq.\,(\ref{eq:sigma_gap_gas}).
Finally, since the gas accretion rate onto the planet cannot be larger than the gas accretion rate onto the central star \citep{lubow_accretion_2006}, the actual gas accretion rate at each time step is given by
\begin{equation}
    \label{eq:gas_acc_rate}
   \Dot{M}_{\mathrm{gas}} = \min [\Dot{M}_{\mathrm{KH}}, \Dot{M}_{\mathrm{disc}}, \Dot{M}_{\star}]
\end{equation}
In our model, we pragmatically cut the gas accretion of our planets once they reach Jupiter mass.

\section{Results}
\label{sect:results}

The following sections are devoted to discuss the impact of a single parameter in our model.
First, we explore the role played by fragmentation velocity, which sets the dimension of the accreted pebbles (Section \ref{subsect:role_vfrag}).
Then, we investigate the role of accretion heating, which modifies the disc aspect ratio and thus influences the pebble accretion rates (Section \ref{subsect:role_visc}). We focus on the role of pebble filtering due to multiple outer giant planets in Section \ref{subsect:role_multiple_giants}.
We consider the possible role of the sublimation of water ice across the ice line (Section \ref{subsect:role_iceline}).
Finally, we explore the consequences of dust leaking through the planetary gap in Section \ref{sec:leaking_dust}.

\begin{figure*}[t!]
\centering
    \includegraphics[width=0.8\textwidth]{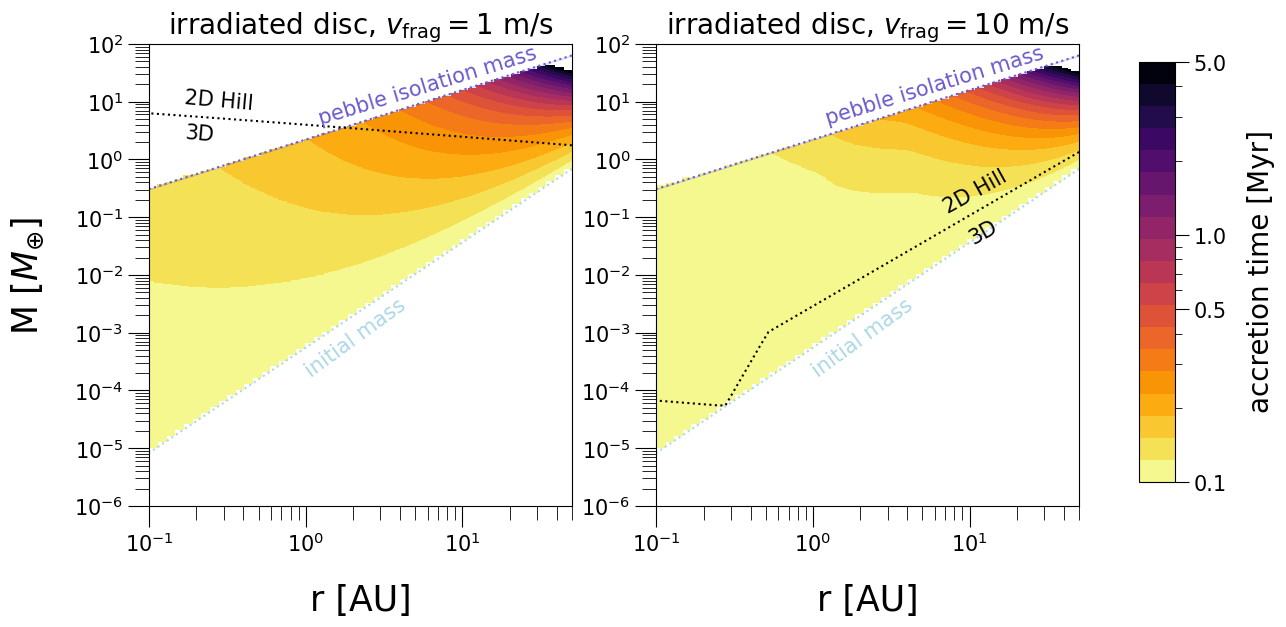}
    \caption{Planetary growth timescales in an irradiated disc, for different fragmentation velocities ($v_{\rm frag}=1 \: \rm m/s$ in the left panel, $v_{\rm frag}=10 \: \rm m/s$ in the right panel). The black dotted line represents the transition mass between the 3D and 2D Hill accretion regimes given by Eq.\,(\ref{eqz:M_3D_2D_trans_reg}). In the high fragmentation velocity case, the transition line is segmented because of the pebble size transition from being drift limited in the Epstein regime to drift limited in the Stokes regime to fragmentation limited (see orange branch, right panel of Fig.\,\ref{fig:st_frag}), while in case of lower fragmentation velocity, the pebbles are always fragmentation limited (blue branch, Fig.\,\ref{fig:st_frag}). 
    The light blue dotted line and the violet dotted line mark the initial embryo masses and the pebble isolation masses, respectively.
    Higher fragmentation velocities lead to faster growth rates, owed to an earlier transition between the 3D and the more efficient 2D Hill accretion regimes. Indeed, the larger pebbles (cfr. Fig.\,\ref{fig:st_frag}) are more settled towards the disc midplane (cfr. Eq.\,\ref{eq:H_peb}), which implies that the criterion in Eq.\,(\ref{eqz:2d3d_trans}) is fulfilled earlier on than in the lower fragmentation velocity case.}
    \label{fig:irr_vfrag}
\end{figure*}

\subsection{The role of fragmentation velocity}
\label{subsect:role_vfrag}
The size of the pebbles is an important parameter for planet formation, because it regulates the degree to which pebbles settle onto the disc midplane \citep{Youdin_2007} and the pebble accretion cross section \citep{ormel_effect_2010,lambrechts_rapid_2012}.
Pebble sizes are limited either by drift or by fragmentation (cfr. Section \ref{subsect:pebble_size}), with a critical fragmentation velocity that depends on their composition. 
Rocky silicate particles have a well-defined fragmentation velocity of approximately $1 \: \rm m/s$ \citep{guttler_outcome_2010}, assuming standard monomers in the $1$\,$\mu$m range \citep{blum_growth_2008}.
If the aggregates grow larger, the fragmentation velocities decrease \citep{Bukhari_2017}, with particles of dm size having fragmentation velocities towards $0.1 \: \rm m/s$. \citep{Deckers_2013}.
Some authors have argued for potentially higher fragmentation velocities beyond $10 \: \rm m/s$, based on the assumption of smaller monomers with $0.1$\,$\mu$m sizes \citep{Kimura_2015} and larger surface energies of silicates \citep{yamamoto_examination_2014}. 

In contrast, icy particles have long been considered to have significantly higher fragmentation velocities, due to their ten times higher surface energies \citep{wada_numerical_2008, gundlach_micrometer-sized_2011}.
However, recent laboratory experiments have shown that water ice and silicate particles have comparable surface energies \citep{Musiolik_2019,Schrapler_2022}, leading to fragmentation velocities of around $1 \: \rm m/s$, possibly increasing closer to the water iceline \citep{Musiolik_2019}. 

To understand the impact of $v_{\rm frag}$ on the efficiency of planet formation, we plot in Fig.\,\ref{fig:irr_vfrag} the accretion time of growing planets in an irradiated disc (\texttt{mod:irrad}), where we ignore planetary migration for clarity.
The accretion time is defined as the amount of time that it takes to an embryo of mass $M_0$ (given by Eq.\,\ref{eq:M0_pla}) to reach a certain mass $M$, while accreting from a pebble flux given by Eq.\,(\ref{eq:M0_pla}), with an accretion efficiency given by Eq.\,(\ref{eq:ff}). 
We use the nominal model parameters as listed in Table \ref{tab:params}.
The left panel shows the accretion time for a fragmentation velocity of $1 \: \rm m/s$, while the right panel refers to  $v_{\rm frag} = 10 \: \rm  m/s$.
The black dotted line marks the transition between 3D and 2D Hill accretion regime at final time ($t=5 \: \rm Myr$). In the high fragmentation velocity case (right panel of Fig.\,\ref{fig:irr_vfrag}) the line appears to be segmented because of the pebble size transition from being drift-limited in the Epstein regime to drift-limited in the Stokes regime, and finally fragmentation-limited (cfr. orange branch, right panel Fig.\,\ref{fig:st_frag}). In the low fragmentation velocity case (left panel of Fig.\,\ref{fig:irr_vfrag}), the transition line is smooth because, within the considered radial extent, the pebbles are always fragmentation-limited (cfr. blue branch, right panel of Fig.\,\ref{fig:st_frag}).

A higher fragmentation velocity leads to an earlier transition between the 3D and the more efficient 2D Hill accretion regimes (black dotted line in Fig.\,\ref{fig:irr_vfrag}) and therefore a faster growth. This is due to the pebbles growing larger for higher $v_{\rm frag}$ (cfr. Fig.\,\ref{fig:st_frag}), thus being more settled towards the disc midplane (cfr. Eq.\,\ref{eq:H_peb}).
By using fragmentation-limited Stokes number (Eq.\,\ref{eq:St_frag}) into Eq.\,(\ref{eqz:M_3D_2D_trans_reg}), we recover, indeed, that the transition mass scales as $M_{\mathrm{3D \rightarrow 2D, Hill}} \propto v_\mathrm{frag}^{-5}$. 

Although increasing fragmentation velocity enhances accretion efficiency in the 3D regime and reduces it in the 2D regime ($f_{\rm 3D} \propto v_{\rm frag} \alpha_z^{-1}$, $f_{\rm 2D, Hill} \propto v_{\rm frag}^{-2/3}$), the 2D accretion efficiency remains higher than the 3D efficiency for both values of $v_{\rm frag}$.
This is because in the 2D regime, the pebbles are accreted from the entire pebble scale height, as opposed to only from a fraction of it in the 3D regime (see the $\alpha_z$ term in the 3D accretion efficiency Eq.\,\ref{eq:acc_3D_eff}, that stems from the pebble scale height Eq.\,\ref{eq:H_peb}), which results in significantly faster embryo growth at higher fragmentation velocities.

For outer planets ($\gtrsim$$5$ AU) with masses above approximately 1\,M$_{\oplus}$
(top right corner of Fig.\,\ref{fig:irr_vfrag}), there is little difference in the growth timescales
between the $1 \: \rm m/s$ and $10 \: \rm m/s$ fragmentation velocity cases, because at those distances pebbles are drift-limited rather than fragmentation-limited (cfr. Fig.\,\ref{fig:st_frag}), rendering the accretion efficiency $v_{\mathrm{frag}}$ independent.

We conclude that in the outer irradiated disc, core growth is only weakly dependent on the chosen fragmentation velocity. In the inner disc, the choice of pebble fragmentation velocity is more relevant, with a higher $v_{\rm frag}$ leading to a faster growth. According to literature, the fragmentation velocities for silicate-dominated pebbles within the iceline are likely around $1 \: \rm  m/s$. Therefore, in our simulations, we proceed with a nominal fragmentation velocity of $v_{\rm frag}=1 \: \rm m/s$.

\subsection{The role of accretion heating}
\label{subsect:role_visc}

\begin{figure*}
    \centering
    \includegraphics[width=\textwidth ]{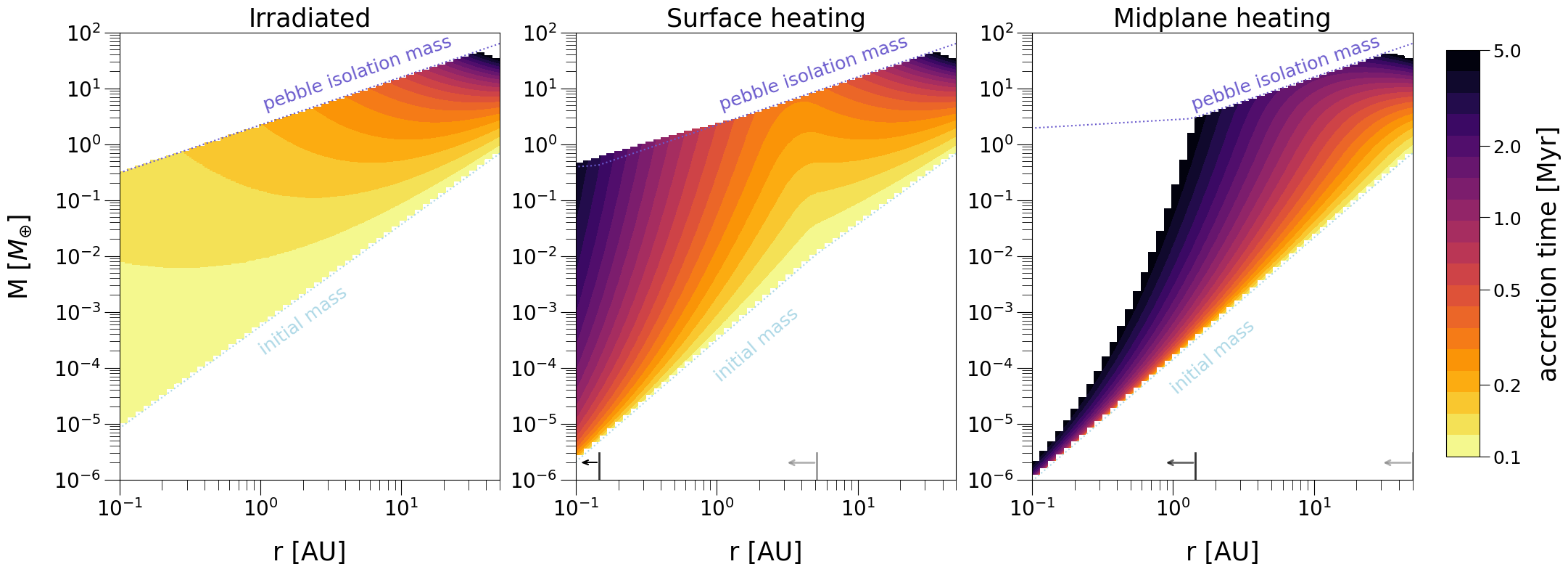}
    \caption{
    Planetary growth timescales of in-situ growing planets for three different disc models: an irradiated disc (left), a surface-heated disc (centre), and a midplane-heated disc (right). The light blue line represents the starting planetesimal mass, while the violet line marks the pebble isolation mass. The small arrows near the x-axis in the central and right panel show the transition radii between the viscous heated inner disc and the irradiated outer disc, respectively at initial time (grey) and at final time (black). Moderate surface heating delays the embryo growth in the inner disc, changing the inside-out growth of a typical irradiated disc into a growth mode that favours the region between $1$ and $10$ AU. Strong midplane heating significantly delays inner growth, suppressing it inside approximately $2$ AU. 
    }
    \label{fig:t_visc}
\end{figure*}

Accretion heating in the inner disc is a fundamental mechanism that sets the temperature profile of the disc.
This, in turn, affects the pebble accretion efficiency (cfr. Eqs. \ref{eq:f_3D}, \ref{eq:f_2DH}) through the Stokes number of the pebbles.  
To investigate the effect of viscous heating on the accretion efficiency of planets, we plot in Fig.\,\ref{fig:t_visc} the accretion time for in-situ growing planets, for three disc models, keeping the nominal $v_{\mathrm{frag}} = 1 \: \rm m/s$.
For comparison, we plot in the left panel the results for the irradiated disc (\texttt{mod:irrad}) which correspond to the left panel of Fig.\,\ref{fig:irr_vfrag}.
The central and right panel represent two discs with moderate (\texttt{mod:surfheat}) and strong (\texttt{mod:midheat}) viscous heating respectively. The difference between the latter two discs lies in the degree of viscous heating expressed through the parameters $\epsilon_{\mathrm{el}}$ and $\epsilon_{\mathrm{heat}}$.
These are set to their nominal values $\epsilon_{\mathrm{el}} = 10^{-2}$ and $\epsilon_{\mathrm{heat}}=0.5$ for the surface-heated disc, where the accretion heating layer is on the disc surface. For a standard midplane-heated viscous disc, where the accretion heating layer is in the midplane, they are set to unity $\epsilon_{\mathrm{el}} = 1$, $\epsilon_{\mathrm{heat}}=1$ (see Table \ref{tab:params} for nominal parameters). The light grey and black arrows on the x-axes of Fig.\,\ref{fig:t_visc} mark the position of the transition between irradiation and viscous heating domain in the disc at the initial and final time of the simulation, respectively.

The more the viscous heating becomes dominant (centre to right panel), the more the growth in the inner disc gets delayed, up to the point of getting suppressed inside approximately 3 AU in case of strong midplane heating.
Therefore, in viscoulsy heated discs (\texttt{mod:surfheat}, \texttt{mod:midheat}, centre and right panel of Fig.\,\ref{fig:t_visc}), core growth tends to be favoured between $1$ and $10$ AU in the former case and between $10$ and $20$ AU in the latter, in contrast to the typical inside-out growth of an irradiated disc (\texttt{mod:irrad}, left panel).
This can be understood by looking at the radial dependence of the pebble accretion efficiency (Eq.\,\ref{eq:ff}), which is the ratio between the amount of material accreted by the planet and the available material set by the pebble flux.
In the 3D accretion regime, the accretion efficiency can be written as
\begin{align}
\label{eq:f_3D}
  f_{\rm 3D} 
  &= 
  \frac{1}{\sqrt{2\pi}}    \left( \frac{d\ln P}{d\ln r} \right)^{-1} 
  {\alpha_z}^{-1/2}{\rm St}^{1/2}
  \left( \frac{H}{r} \right)^{-3}
  \left(\frac{M}{M_{\star}}\right) \, .
\end{align}
Using fragmentation-limited pebbles, which is an approximation that holds in the inner disc, we can rewrite
\begin{align}
\label{eq:acc_3D_eff}
  f_{\rm 3D,frag} 
   \propto v_{\rm frag} \alpha_{\rm frag}^{-1/2} \alpha_z^{-1/2} \left(\frac{M}{M_{\star}}\right)
  \left( \frac{H}{r} \right)^{-4}
  r^{1/2} \,.
\end{align}
This expression allows us to understand how the pebble accretion efficiency behaves as a function of radius. 
In a viscous disc, the aspect ratio given by Eq.\,(\ref{eq:H_R_visc}), increases positively with radius as $H/r \propto r^{1/20}$.
This means that the accretion efficiency increases at wider orbits, as $f_{\rm 3D, frag} \propto r^{3/10}$, suppressing growth close to the host star.
In contrast, in an irradiated disc, the the disc aspect ratio increases with radius as $H/r \propto r^{2/7}$ (Eq.\,\ref{eq:H_R_irr}). 
Therefore, the accretion efficiency decreases at wider orbits, as $f_{\rm 3D, frag} \propto r^{-9/14}$, quenching planet formation outside of 50 AU.

We can do the same analysis for the 2D Hill accretion regime
\begin{align}
    \label{eq:f_2DH}
    f_{\mathrm{2D \: Hill}} 
    &= 
     \left(\frac{6}{\pi^3}\right)^{1/3} \left( \frac{d\ln P}{d\ln r} \right)^{-1} \mathrm{St}^{-1/3} \left(\frac{M}{M_{\star}}\right)^{2/3}  \left( \frac{H}{r} \right)^{-2}. 
\end{align}
The accretion efficiency for fragmentation-limited pebbles then scales as
\begin{gather}
    \label{eq:f_2DH_frag}
  f_{\rm 2D \: Hill, frag} \propto  \alpha_{\mathrm{frag}}^{1/3} v^{-2/3}_{\mathrm{frag}} \left(\frac{M}{M_{\star}}\right)^{2/3}  \left( \frac{H}{r} \right)^{-4/3} r^{-1/3}.
\end{gather}
In this case, the viscous disc will lead to an accretion efficiency of $f_{\rm 2D \: Hill, frag} \propto r^{-4/15}$, while the irradiated disc gives $f_{\rm 2D \: Hill, frag} \propto r^{-5/7}$. Thus, in the case of 2D Hill accretion, the radial dependence is negative in both cases, but scales almost three times as steep for the irradiated disc. This leads again to a more efficient accretion rate in the inner disc in the irradiated case, compared to the viscously heated case. 

The above discussion assumes that pebble accretion occurs in the so-called slow encounter regime \citep[see][ for a review]{Ormel_emerging_paradigm_2017}, where the encounter timescale is sufficiently long for the pebble to efficiently settle onto the planetary embryo. However, this condition is not necessarily satisfied around small planetesimals in the inner disc, where the pebble approach velocity ($\Delta v$) could exceed a critical velocity $v_{\rm crit}$ causing inefficient settling and reducing the pebble accretion efficiency \citep{Liu_2018, ormel_2018}. We verified that this condition is not met in our disc models and for the initial planetesimal masses considered here. Poor pebble settling and reduced pebble accretion efficiency would become important only in the strongly viscous case (\texttt{mod:midheat}) for small planetsimals ($\lesssim  10^{-6} M_{\oplus}$) inside 0.1 AU, a region not considered here.

In summary, for fragmentation-limited pebbles, with fragmentation velocities around $1 \: \rm m/s$, the degree of accretion heating in the inner disc is important. Efficient heat deposition near the disc midplane raises the gas scale height and reduces the degree of disc flaring.
This breaks the strong inside-out growth trend present in purely irradiated discs (Fig.\,\ref{fig:t_visc}), proving that the choice of disc model is key in determining the growth of planetary embryos.
Since the degree of accretion heating is poorly constrained, this model uncertainty has to be kept in mind going forward.

\subsection{The role of outer giant planets}
\label{subsect:role_multiple_giants}

\begin{figure*}[ht!]
    \centering
    \includegraphics[width=\linewidth]{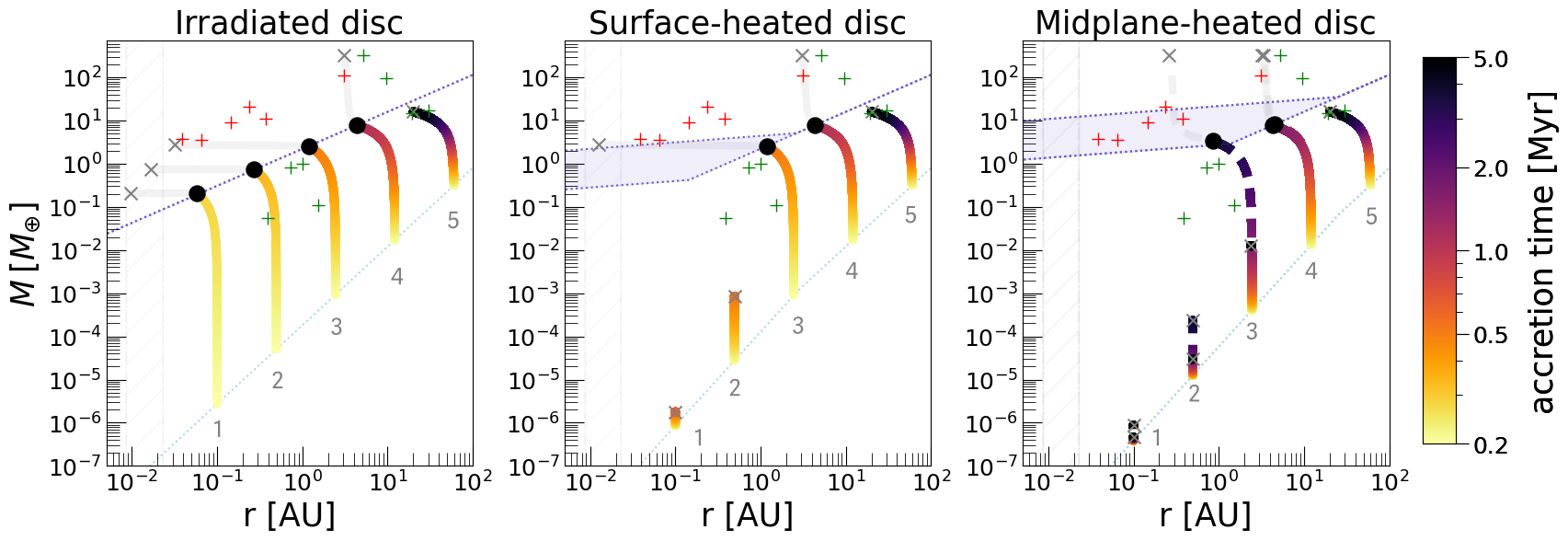}
    \caption{
    Growth tracks of a system of 5 planets for an irradiated (left panel), surface-heated (central panel) and midplane-heated (right panel) disc, the solid lines represent a system with mutual filtering, while the dashed lines in the right panel show the same system but without mutual filtering. Colour-coded is the accretion timescale.
    The dotted light blue line is the initial mass of the embryo distribution. The violet dotted line represents the pebble isolation mass, which is time dependent for the viscously heated discs, therefore the violet shaded region shows the changing location of the isolation mass between initial and final time of the simulation.
    The grey shaded area marks the magnetospheric cavity radius, which shifts outwards with time. 
    The black dots identify where the planet crosses the pebble isolation mass, while the grey crosses represent the final mass and position of the planets.
    The grey lines mark the gas accretion phase.
    For reference, the green and black crosses represent the Solar System and the HD 219134 planets \citep{Vogt_2015} respectively.
    Mutual filtering is irrelevant in irradiated discs because of the fast inside-out growth mode, while it can become relevant for strong midplane heating, aided by the growth delay in the inner disc due to viscous heating.
    }
    \label{fig:multiple_filter_nofilter}
\end{figure*}

We now simulate systems with multiple embryos growing simultaneously to study how the growth of inner embryos is affected by outer embryos accreting a fraction of the available pebble flux.
The planetary embryos are distributed with a uniform logarithmic spacing and have an initial mass given by Eq.\,\ref{eq:M0_pla}.
They accrete in the different pebble accretion regimes (3D/2D, Bondi/Hill) according to their growing mass, as explained in 
Section \ref{subsect:pebble_accretion}.
Each embryo filters a fraction of the pebble flux as in Eq.\,(\ref{eq:ff}), so that the embryos inside its orbit accrete from a reduced flux (as in Eq.\,\ref{eq:F_p}). All embryos also undergo type I migration (Eq.\,\ref{eq:typeI_mig}) as they grow, until they reach pebble isolation mass, after which they switch to gas accretion (see Section \ref{subsect:gas_accretion}) and type II migration (Eq.\,\ref{eq:typeII_mig}). 
Table \ref{tab:params} summarises the nominal values of our simulations.

We compare the growth tracks of a system of 5 planets distributed logarithmically between 0.1 and 60 AU for an irradiated disc (left panel), a surface-heated disc (central panel) and a midplane-heated disc (right panel) in Fig.\,\ref{fig:multiple_filter_nofilter}. 
We used a fixed fragmentation velocity of $1 \: \rm m/s$. 
The colour-coding of the growth track shows the accretion time.
The dotted light blue and violet lines mark the initial masses and pebble isolation mass respectively. In the viscously heated scenarios (central and left panel of Fig.\,\ref{fig:multiple_filter_nofilter}), the pebble isolation mass changes with time due to the time dependency in the disc aspect ratio. The violet shaded area marks the time evolution of the pebble isolation mass from $t_{\rm in} = 0.2$ Myr to $t_{\rm fin} = 5$ Myr.
The black dots mark the planets' isolation mass, while the grey crosses mark the final mass and position of the planets. If the final masses are below the isolation mass, it means that either the planets did not manage to grow to isolation mass within the disc lifetime, or that they stopped growing due to complete pebble filtering from outer planets. Numbers identifying each planet are placed to aid the description of the figure.

In the background we also show, as reference, with blue crosses the Solar System and with green crosses a representative exoplanet system with inner super-Earths and outer cold giants (See also Section\,\ref{subsec:cond}).
We selected the HD 219134 exoplanet system from the catalogue of known super-Earth systems with outer giant companions presented in \citet{van_zandt_2025}, where they follow up TESS detections of super-Earths with a Keck RV survey.
HD 219134  stands out as a well-characterized system \citep{Vogt_2015} with 5 detected super-Earths and one giant planet outside of $\approx 2.5$ AU.

In the irradiated disc (left panel of Fig.\,\ref{fig:multiple_filter_nofilter}), the filtering of the outer planets on the inner embryos is irrelevant.
Inside-out growth allows the inner embryos to grow to isolation mass before the outer embryos manage to filter a significant fraction of pebbles.
This leads to a final system where the four innermost planets all reach isolation mass and switch to gas accretion, giving rise to super-Earths and gas giants, while the outermost embryo stalls at the ice giant stage due to the much longer accretion time required in the outer disc (cfr. Fig.\,\ref{fig:irr_vfrag} top right corner).
We halt the migration of the inner planets when they reach the magnetospheric cavity radius (Section \ref{subsect:typeI_mig}). Here, we ignore the possibility that planets may be pushed outwards as the cavity expands with time \citep{liu_dynamical_2017}.  
To avoid all planets to pile up at exactly the inner edge location, we pragmatically prevent planets with a mass beyond the pebble isolation mass to come within a 2:1 period ratio from each other. This mimics the common process of convergent migration in the type-I migration regime, where planets get trapped in resonant chains with 2:1 or 3:2 period ratios \citep{Terquem_2007,Kajtazi_2023}.  

In the surface-heated disc (central panel, Fig.\,\ref{fig:multiple_filter_nofilter}), the delayed growth in the inner disc allows the third planet to reach isolation mass before the inner embryos, halting the pebble flux and migrating all the way to the edge, preventing their growth.
In this case, indeed, we see convergent migration, with a more massive outer embryos migrating towards lower-mass inner planets. As we do not model planet-planet gravitational interactions, we pragmatically choose to halt the growth of a small protoplanet when crossed by an outer planet migrating inwards.

In the case of strong midplane-heating (right panel, Fig.\,\ref{fig:multiple_filter_nofilter}), the growth of the 3 innermost embryos is completely suppressed. In this panel we also show the growth tracks of the system when mutual pebble filtering is not considered, by overplotting a dashed line. The growth of the two innermost embryos is suppressed both in the mutual filtering and in the no filtering case (solid and dashed line respectively), implying that it is owed almost entirely to the effect of viscous heating. Mutual filtering plays the biggest difference for the third planet, that in the no filtering case reaches pebble isolation mass and grows into a gas giant (dashed line), while in the mutual filtering case stalls at Mars-mass (solid line). The delay in the inner growth timescale of a midplane heated disc, indeed, allows the fourth embryo to reach isolation mass faster than the inner embryos (1-3), thus cutting the entire pebble flux and preventing accretion (we relax this assumption in Section \ref{sec:leaking_dust}).

We conclude that mutual pebble filtering between outer giants and inner embryos could play a relevant role in suppressing super-Earth formation, only if it is accompanied by a mechanism that delays growth in the inner disc, such as strong midplane viscous heating. Without such a mechanism, pebble filtering alone is not effective enough to prevent super-Earth formation.

\subsection{Conditional occurrence of super-Earths and cold giants}
\begin{table}[t!]
\centering
    \caption{Planet type definitions}
    \begin{tabular}{ ccc }
        \hline
        \hline
        Planet type &  Mass & Orbital distance  \\
        \hline
        Hot Jupiters & $100 < M_{\rm p}< 6000 \: M_{\oplus}$ & $a_{\rm p} < 0.1 \: \rm AU$ \\
        Warm Jupiters & $100 < M_{\rm p}< 6000 \: M_{\oplus}$ & $0.1<a_{\rm p} < 2 \: \rm AU$ \\
        Super-Earths & $1 < M_{\rm p}< 20 \: M_{\oplus}$ &  $a_{\rm p} < 1 \: \rm AU$ \\
        Sub-Earths & $0.01 < M_{\rm p}< 1 \: M_{\oplus}$ &  $0.1<a_{\rm p} < 10 \: \rm AU$ \\

        \hline
    \label{tab:planets_def}

    \end{tabular}
\end{table}

\begin{figure*}
    \centering
    \includegraphics[width=\textwidth]{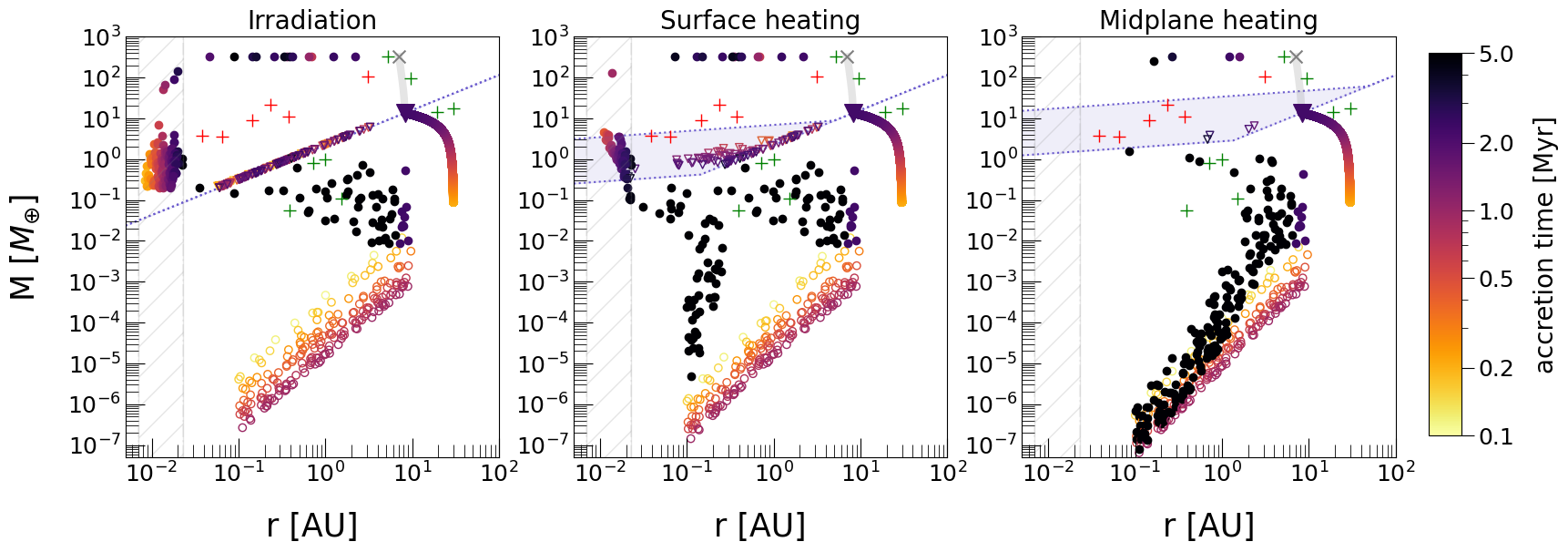}
    \caption{Population synthesis of pairs of planets (Jupiter-like planet + one inner embryo) for the three different disc models considered in this study. The empty dots represent the random initial positions and masses of the inner embryos, the triangles mark the position at which the embryos cross isolation mass and the full dots represent the final position and masses of the planets. All markers are colour-coded by accretion time. The position and start time of the Jupiter-like planet is fixed for each simulation at $a_{\mathrm{p,0}} = 30$ AU and $t_0 = 2.5 \cdot 10^{5}$ yrs. The filtering due to the Jupiter-like planet is basically irrelevant in the irradiated disc as almost all embryos reach isolation mass leading to a population of super-Earths and some gas giants. In a moderately viscous disc (central panel) growth is partially hindered by the combined effect of delayed growth of inner embryos and pebble filtering, while a strongly viscous heated disc (right panel) results in completely hindered growth in the inner disc and Mars-mass embryos in the outer disc. For reference, the green and black crosses represent the Solar System and the HD219134 planets respectively.
}
    \label{fig:param_explor} 
\end{figure*}

To explore the dependency on the time and location where the inner embryos emerge, we perform a set of 200 simulations drawing an initial embryo distribution in a population synthesis-like fashion.
Each simulation features the same Jupiter-like planet with initial position $a_{\rm out, 0} = 30$ AU and initial mass $M_{\rm out, 0} \simeq 0.088 M_{\oplus}$ and one inner embryo, whose initial time and position are randomly drawn from a linear uniform distribution in time $0.1<t<1 \rm \,Myr $ and a log-uniform distribution in space $0.1< a_{\rm in,0}< 10 \rm \,AU$. 
The results are shown in Fig.\,\ref{fig:param_explor}, where the empty dots mark the initial position and mass of the inner embryo and are colour coded by time of insertion of the embryo in the disc. The triangles mark the position at which the embryos reach pebble isolation mass, while the full dots show the final positions and masses of the embryos, both also colour-coded by accretion time.
For clarity, the full growth track is only shown for the outer Jupiter-like planet.
The different panels in Fig.\,\ref{fig:param_explor} show the same set of simulations but for the three different disc models.
Table \ref{tab:planets_def} summarises our planetary types definitions.

In the irradiated disc (left panel of Fig.\,\ref{fig:param_explor}), the final outcome of the ``population synthesis'' is a small occurrence of hot ($ < 2 \%$) and warm gas giants ($\approx 5 \%$), a set of super-Earths in the inner disc ($\approx 20 \%$), and a similar fraction of planets ($\approx 25 \%$) smaller than Earth ($0.01<M<1 \: M_{\oplus}$) that orbit farer out ($0.1< a_{\rm p}< 10 \: \rm AU$), which we will refer to from now on as sub-Earths.
These were the outermost embryos with growth timescales comparable to the Jupiter-like planet, whose growth has been hindered by mutual filtering. 

The surface-heated disc model (central panel of Fig.\,\ref{fig:param_explor}) confirms the trend already seen in Fig.\,\ref{fig:multiple_filter_nofilter}. 
The innermost embryo's growth is halted by the effect of pebble filtering aided by the inner growth's delay (cfr. central panel of Fig.\,\ref{fig:t_visc}).
The number of formed super-Earths in this case is $\approx 25 \%$ but some are slightly less massive.
The amount of sub-Earths is consistent with the irradiated disc ($\approx 27 \%$), as most form in the outer part of the disc which is always irradiation dominated and subsequently migrate inwards.

As already hinted by Fig.\,\ref{fig:multiple_filter_nofilter}, the midplane-heated disc (right panel of Fig.\,\ref{fig:param_explor}) features very few planets that reach isolation mass.
Super-Earth formation is almost entirely suppressed ($< 2 \%$), while the fraction of sub-Earths is slightly reduced but still comparable to the other disc models ($\approx 22 \%$), with the difference that in this case the sub-Earths are predominantly distributed outside $1$ AU orbits. This is consistent with the accretion timescale in Fig.\,\ref{fig:t_visc}, showing that growth is more efficient in a preferred region outside 5 AU and completely hindered inside $2$ AU. 

Summarising, we find that, in general, the pebble flux needed to form the cores of cold giants is enough to form systems of super-Earths in closer orbits. 
Only when the inner disc is sufficiently accretion heated ($\epsilon_{\rm el} \gtrsim 10^{-2}$), the inner embryo growth is delayed to such a degree that pebble filtering by outer cores outside the iceline becomes an important additional factor in suppressing inner-disc core growth.

\subsection{The role of the iceline}
\label{subsect:role_iceline}

We define the radial position of the iceline as the location in the disc where the temperature reaches $T_{\rm H_{2}O} = 170$ K. In case of an irradiated disc, the iceline is at a fixed location
\begin{equation}
    \label{eq:iceline_irr}
    r_{\mathrm{H_2O, irr}} \simeq 0.688 \left(\frac{L_{\star}}{L_{\odot}}\right)^{2/3} \left(\frac{M_{\star}}{M_{\odot}}\right)^{-1/3}  \rm{AU}\,.
\end{equation}
In a viscously heated disc, the location of the iceline inherits the gas accretion rate time dependency (Eq.\,\ref{eq:sigma-gas_dot}), meaning that the iceline moves inwards with time as the disc cools down
\begin{align}
  r_{\mathrm{H_2O, visc}} 
     & \simeq  0.498
     \left( \frac{\epsilon_{\rm el}}{10^{-2}} \right)^{2/9}
     \left( \frac{\epsilon_{\rm heat}}{0.5} \right)^{2/9}
     \left( \frac{\alpha_{\nu}}{10^{-2}} \right)^{-2/9}
     \notag\\
     &
    \left(\frac{Z}{0.01} \right)^{2/9} 
     \left( \frac{a_{\rm gr}}{0.1\,{\rm mm}} \right)^{-2/9}
    \left(\frac{\rho_{\rm gr}}{1\,{\rm g\,/cm}^3} \right)^{-2/9}   
    \notag\\
    &
    \left(\frac{M_{\star}}{M_{\odot}}\right)^{1/3} 
     \left( \frac{\dot M_{\star}}{10^{-8} M_\odot{\rm /yr}} \right)^{4/9} \rm{AU},
\end{align}
going from $\approx 1.9 \rm \: AU$ to $\approx 0.3 \rm \: AU$ in the surface-heated disc and moving between $\approx 6.3 \rm  \: AU$ to $\approx 0.9 \rm \: AU$ in the midplane-heated case.
Here we investigate the role of the iceline considering three possible effects.
We first explore a reduction in pebble flux by a factor 2 across it, by setting $f_{\rm iceline} = 0.5$ in equation (\ref{eq:F_0}) inside the iceline. This is to mimic 
volatile loss through water sublimation as pebbles cross the iceline. A reduction by a factor 2 is based on a solar mixture \citep{Lodders_2003}, but may be too large given that the refractory-to-ice mass ratios in comets is estimated to be above 3 \citep[see ][ review on comet 67P/Churyumov-Gerasimenko]{Choukroun_2023}. However, a new approach by \citet{Marschall_2025} estimated the refractory-to-ice ratio in 67P to be in between 0.5-1.7, showing that the estimate is still under debate, thus we proceed with the factor 2 for our analysis.

We also explore the possibility of an increase in turbulent stirring ($\alpha_z$) inside the water ice line. This is mainly to explore the effect of this parameter, but recent observations do, tenetavily, indicate that the inner disc might be more turbulent 
compared to the outer disc \citep{jiang_grain-size_2024}.

In the last step, we also include the option of increasing the fragmentation alpha ($\alpha_{\rm frag}$) 
to be aligned with the level of turbulent stirring. 
In this case, the pebble size decreases when crossing the water iceline (cfr. Eq.\,\ref{eq:St_frag}), which is believed to occur if the fragmentation velocity of icy pebbles differs from that of bare silicates (see Section \ref{subsect:role_vfrag}), or if pebbles fall apart upon water ice sublimation \citep[many-seeds model, see][]{Aumatell_2011, Schoonenberg_2017, Houge_2023}. Note that the latter scenario seems unlikely, as pebbles have been found to resist complete destruction even upon intense heating 
events \citep{Houge_2024}.

\begin{figure*}
    \centering
    \includegraphics[width=\textwidth]{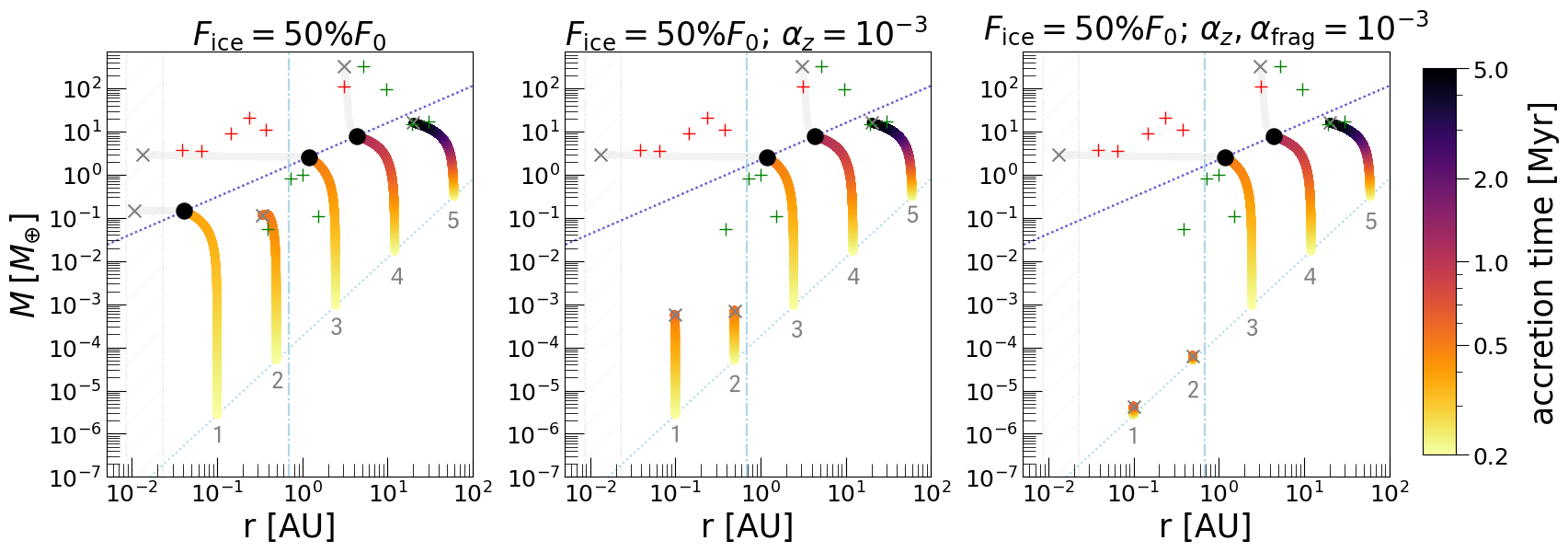}
    \caption{Possible role of the iceline in suppressing embryo growth in the inner disc.
    Growth tracks of a system of 5 planets logarithmically spaced between 0.1 and 60 AU in an irradiated disc, with colour-coded growth timescale. The left panel shows the effects of a $50 \%$ flux reduction across the iceline, the central panel shows the cumulative effect of flux reduction and increase in the vertical stirring of pebbles and finally the right panel shows the cumulative effect of flux reduction and increase in both vertical stirring and $\alpha_{\rm frag}$. 
    The vertical light-blue line marks the position of the iceline, the grey dashed area the expanding inner magnetospheric cavity radius. The pebble isolation mass is indicated with a purple dotted line. 
    The black dots mark the position at which the planets reach isolation mass, while the grey lines and crosses show the planet's gas accretion and final mass and position respectively.
    The pebble flux reduction alone is not sufficient to prevent inner embryo's growth. Only by combining the effect with an increased vertical stirring and/or fragmentation turbulence the inner growth gets completely suppressed.
        }
    \label{fig:iceline_eff}
\end{figure*}

We show the cumulative effects of these processes in Fig.\,\ref{fig:iceline_eff}, using the same system of 5 planets as in Fig.\,\ref{fig:multiple_filter_nofilter} and an irradiated disc to have maximal inner growth efficiency.
The left panel shows a $50\%$ flux reduction across the iceline, the central panel shows the cumulative effect of the flux reduction and the increase in vertical turbulence, and finally the right panel shows the cumulative effect of flux reduction and both $\alpha_z$ and $\alpha_{\rm frag}$ increase.

Comparing the left panel of Fig.\,\ref{fig:iceline_eff} with the left panel of Fig.\,\ref{fig:multiple_filter_nofilter}, we see that a $50 \%$ flux reduction only affects the second innermost planet, preventing it from reaching pebble isolation mass. This is due to the fact that the flux reduction slows down the inner embryos' growth enough that the third innermost planet reaches pebble isolation mass and switches to gas accretion before the second innermost planet, halting its growth. The innermost planet 
grows still fast enough to reach isolation mass and 
undergoes gas accretion before complete pebble filtering, due to the inside-out growth mode. Thus, merely reducing the pebble flux across the iceline is not enough to prevent super-Earth formation in the inner disc.

The central panel of Fig.\,\ref{fig:iceline_eff} shows the combined effect of reducing the flux and increasing the vertical turbulence of the inner disc. This results in a suppression of the growth of the inner embryos. Increasing $\alpha_z$ means increasing the pebble scale height, thus decreasing the accretion rate in the 3D regime, effectively slowing down accretion. This, in turn, implies that the outer planets (3,4,5) reach isolation mass first, completely halting the pebble flux and damping inner disc growth. 

The right panel, finally, shows the cumulative contribution of the pebble flux reduction and the increase in both the vertical turbulence and $\alpha_{\rm frag}$. The inner embryos (1,2) in this case barely manage to grow. Increasing $\alpha_{\rm frag}$ results in smaller pebble sizes, thus less efficient 3D accretion (cfr. Eq.\,\ref{eq:f_3D}), adding another obstacle to the already hindered inner disc planet formation.
This allows the third planet to reach isolation mass and completely prevent any type of inner growth.
This is consistent with what is shown in Fig. 2 of \citet{mulders_why_2021}, where they show that the filtering due to a Jupiter-like planet is able to prevent super-Earth formation if the iceline features a $50 \%$ reduction of the pebble flux and a factor 10 reduction of the fragmentation velocity. Although we increase $\alpha_{\rm frag}$ and do not change $v_{\rm frag}$, the final effect is always a reduction in the Stokes number of the pebble (see Eq.\,\ref{eq:St_frag}).

In summary, even if pebbles would loose half of their mass through water sublimation, embryos still grow efficiently in the inner disc. If, for some reason, turbulent stirring rates are also increased or the particle sizes are reduced when crossing the iceline, the delay in the inner embryos' growth becomes substantial enough to allow pebble filtering to completely prevent super-Earth formation.
It is unclear why this would be the case in some discs and not others. 

\subsection{Efficiency of dust leaking through the planetary gap}
\label{sec:leaking_dust}
Several studies have proven that gaps opened by massive planets do not completely block dust outside their orbit, but are rather permeable, allowing some fraction of particles to leak through. 
Small dust particles, that are well-coupled with the gas, follow the viscous accretion and 
pass through the gap.
The critical size of the dust that gets trapped outside of the gap depends on the disc
viscosity, the planet mass, and the stellar accretion rate \citep{Weber_2018, Haugbolle_2019, van_clepper_2025}. For a disc with $\alpha_{\nu} = 10^{-3}$ and a Jupiter-like core, only $\mu \rm m$ size dust is able to leak, however, for smaller planetary cores ($50 \, M_{\oplus}$) the critical size can be around $a_{\rm gr} \approx 1 \, \rm mm$, increasing for higher disc viscosities. 
Not only can the dust leak through the gap opened by the planet, but \citet{stammler_leaky_2023} has recently shown that particles trapped in the outer edge of the gap rapidly fragment and are then transported through the gap.

\begin{figure*}
    \centering
    \includegraphics[width=\textwidth]{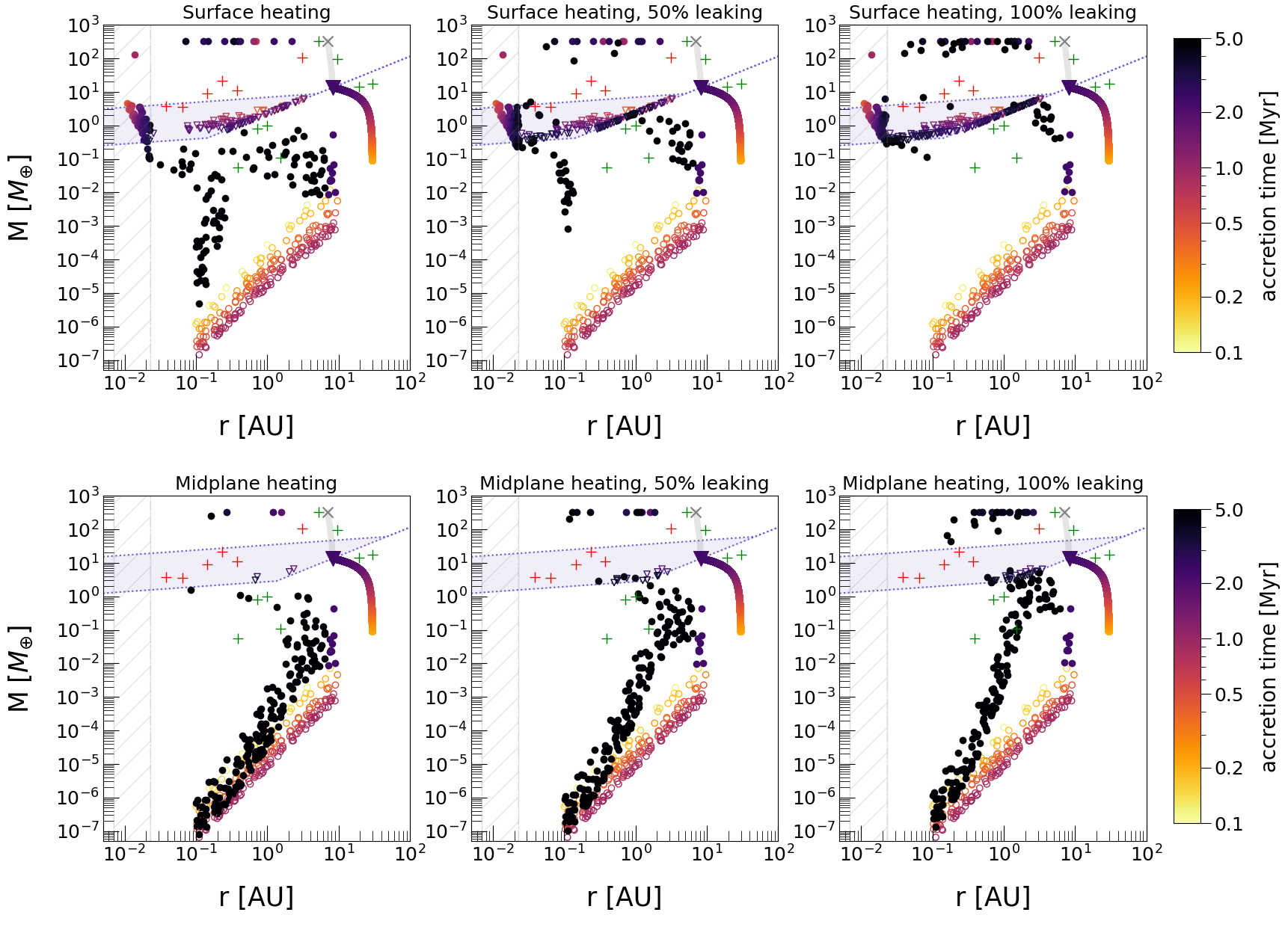}
    \caption{ Possible effect of leaking pebbles through the giant planet's gap. Same set of simulations as in Fig.\,\ref{fig:param_explor} for surface-heated (top row) and midplane-heated (bottom row) discs but with leaking efficiency through the giant planet's gap after isolation mass. The first row shows surface-heated discs, respectively without, with $50 \%$ and with $100 \%$ of pebbles leaking through the gap carved by the Jupiter-like planet after reaching pebble isolation mass. The second row shows the same, but for a midplane-heated disc. A planetary gap that allows for half of the pebble flux to leak through promotes the growth of the outermost embryos in the population in both models.
    The difference lies in the fact that, in the weaker surface-heating case, nearly all the inner embryos grow into super-Earths, while in the midplane-heating case, the embryos grow bigger in mass but still remain firmly below isolation mass.
    A complete permeability of the gap leads to systems of only super-Earths and gas giants in surface-heated discs, while in the miplane-heating case, despite being promoted, the embryo growth is still suppressed inside 1 AU.}
    \label{fig:param_explor_50_0}
\end{figure*}

Based on these recent studies, we investigated the outcome of our simulations in case the planets would not entirely block out the pebble flux once they reach pebble isolation mass. Fig.\,\ref{fig:param_explor_50_0} shows the same set of simulations as Fig.\,\ref{fig:param_explor}, but considering a $50 \%$ leaking efficiency of the pebble flux through the gap (central column) and a full $100 \%$ leaking efficiency (right column). The two different rows show the surface-heated (top) and the midplane-heated (bottom) disc, while the left panels showcase the case of no leaking (same as Fig.\,\ref{fig:param_explor}) for comparison.
Allowing $50 \%$ of the pebble flux to leak through the gap promotes more growth throughout the disc for both models, however the effects in the inner disc are negligible in the strong miplane heating case. This is because the growth suppression inside 1 AU for the midplane heating model is primarily attributed to viscous heating rather than mutual filtering (cfr. Fig \ref{fig:t_visc} and Fig.\,\ref{fig:multiple_filter_nofilter}).
Allowing the entire pebble flux to leak through the gap leads to complete growth for nearly all planets in the surface-heated scenario. In contrast, in the strongly midplane-heated disc, this results in more outer embryos reaching pebble isolation mass and more growth in the inner disc. However, the growth of embryos within 1 AU remains significantly suppressed.

In summary, dust leaking through the gaps of outer giant planets could influence the formation of inner planets. If the leaking efficiency exceeds $50\,\%$, it becomes difficult to imagine how the presence of outer giants could prevent the formation of super-Earths on closer orbits, even in discs with moderate accretion heating.

\section{Discussion}
\label{sec:discussion}

\subsection{Conditional occurrence of super-Earths and cold giants}
\label{subsec:cond}

Our work argues that it is difficult to prevent the formation of super-Earths through pebble accretion in systems with cold giants if the protoplanetary disc is dominated by irradiation or is only weakly accretion-heated.
This finding seems in line with occurrence rate studies, although it remains somewhat unclear to what degree the presence of cold giants influences the occurrence of smaller planets in closer orbits. In this section we briefly review the different constraints reported in the literature.

Observational constraints show either that the presence of a gas giant 
(i) does not significantly affect the occurrence of super-Earths \citep{rosenthal_california_2022, bonomo_cold_2023, Bonomo_2025}, or  that 
(ii) it moderately enhances super-Earth presence \citep{van_zandt_2025}, or that 
(iii) almost certainly implies the presence of at least one super-Earth \citep{zhu_super_2018, bryan_excess_2019}.

Our conditional occurrence rate for irradiated and surface heated disc models averages at $P(\rm SE|CJ) \approx 22.5 \%$, but decreases to $P(\rm SE|CJ) \approx 15 \%$ if we include the midplane heated disc case. Comparing it to the observed field occurrence of super-Earths $P(\rm SE) = 30 \pm 3 \%$ \citep{zhu_about_2018}, we see a slight decrease in the occurrence rate. Our result is still approximately consistent with the conditional occurrence rate between a Jupiter analogue\footnote{\citet{rosenthal_california_2022} define as a Jupiter analogue a planet that orbits $3 \mathrm{AU}<a_{\mathrm{p}}<7 \mathrm{AU}$, with mass in the range $95M_{\oplus}<M_{\mathrm{p}}<4130M_{\oplus}$, and a close-in small planet as a planet orbiting at $0.023 \mathrm{AU}<a_{\mathrm{p}}<1 \mathrm{AU}$, with mass in range $2M_{\oplus}<M_{\mathrm{p}}<30M_{\oplus}$} and close-in small planet $P(\mathrm{I|J}) = 32_{-24}^{+16}\%$ given by \citet{rosenthal_california_2022}, as well as the occurrence rate\footnote{\citet{bryan_friends_2024} define as a gas giant a planet with mass in the range $0.5 \: M_{\rm J} < M< 20 \: M_{\rm J}$, while a super-Earth as a planet with mass in the range $1 \: M_{\oplus}<M<20 \: M_{\oplus}$, and radius in the range $1 \: R_{\oplus}<R<4 \: R_{\oplus}$} provided by \citet{bryan_friends_2024} for solar and sub-solar metallicities  $P(\rm SE|GG, [Fe/H] \leq 0) = 13.0^{+18.9}_{-5.1}\%$.

A complication when interpreting the observed conditional occurrence rates is that they appear to be metallicity dependent. 
\citet{bryan_friends_2024} found a positive correlation for metal-rich stars which disappears for metal-poor stars. Furthermore, \citet{zhu_metallicity_2024} showed that, using only the super-solar metallicity stars in the \citet{bonomo_cold_2023} catalogue, the conditional occurrence rate of cold Jupiters and super-Earths would be consistent with the excess observed by \citet{bryan_excess_2019}.
However, \citet{van_zandt_2025}, in contrast to previous studies, found no evidence that stellar metallicity enhances the conditional occurrence between cold giants and super-Earths.
Based on the models presented here, a stellar metallicity dependency on the conditional occurrence of super-Earths given the presence of cold giants would be weak, because selecting systems with cold giants already fixes a pebble flux sufficient for super-Earth formation, regardless of stellar metallicity \citep{chachan_small_2023}.

\subsection{Disc models}

So far, our study underlines that only a strong midplane accretion heating paired with pebble filtering due to the simultaneous formation of an outer giant planets is able to suppress super-Earth formation, showing that the disc model choice has a big impact on the outcome of the simulations.

To explore the uncertainty in our understanding of the inner accretion disc around young stars, we explored different disc heating models. Nevertheless, several aspects could be explored in more depth. 
For example, the very inner disc,  with temperatures above 800\,K, is likely sufficiently thermally ionized to drive MRI turbulence \citep{Desch_2010}. At wider orbital locations, the weak level of ionization damps the MRI in most regions of the disc \citep{Perez-Becker_2011}, which argues against strong levels of midplane heating. 
Instead, discs evolving through disc winds, with accretion heating in the upper disc surface layers, may be more favoured \citep{Mori_2019}. Although we approximate this process here, we do not explicitly evolve the disc gas surface density through a disc wind torque \citep{Suzuki_2016, Bai_2016}, which is challenging without resorting to non-ideal MHD simulations \citep{Martel_2022, Rea_2024}. Recently, \citet{Batygin_2024} included a MHD wind disc model in their simulation for planet formation.
They argued that pebble accretion in the context of terrestrial planet formation is a dominant process only in discs which are MHD-wind driven with low viscosity.
Their findings are consistent with our parameter exploration: low midplane turbulence ($\alpha_z = 10^{-4}$) is a requirement for efficient pebble accretion. If MRI turbulence penetrates to the disc midplane,  such higher midplane turbulence values ( $\alpha_z \geq 10^{-3}$)  suppress inner disc pebble accretion (cfr. Fig.\,\ref{fig:iceline_eff}). A similar conclusion was drawn by \citet{chachan_small_2023}.

Finally, in order to determine midplane temperatures, we used a simplified constant dust opacity (cfr. Appendix \ref{subsec:opa}). The inner disc opacity, likely provided by fragmentation-generated dust, is highly uncertain, but key in deriving accurate temperatures \citep{Savvidou_2020}.  

\subsection{Stellar mass dependency}
In this paper we have not explored possible stellar-mass dependencies, leaving it to further studies to expand our model and take into account different host star masses to be compared with observations.
Nevertheless, observations show
the occurrence rate of gas giants around M-dwarfs is smaller than around FGK stars \citep[e.g.][]{Clanton_2014,fulton_california_2021}. However, the occurrence of super-Earths peaks around early M-darfs \citep{Mulders_2015, Hsu_2020, Sabotta_2021, Ment_2023, Pan_2025}. This is surprising, if the conditional occurrence rate between super-Earths and cold giants seems to point towards a slightly positive correlation \citep[e.g.][]{zhu_about_2018, van_zandt_2025}.

To reconcile these seemingly contrasting observational results, several theoretical studies have been conducted. 
\citet{mulders_why_2021} argue that, around FGK stars, giant planets cores that reach pebble isolation mass trigger runaway gas accretion and cut the pebble flux, thereby preventing inner super-Earth formation. In contrast, around M-dwarfs only planet cores in the most massive discs reach isolation mass and can become gas giants. For most of the discs, however, conditions for gas giant planet formation are not met, therefore, without this potential reduction of the pebble flux, super-Earths can form more efficiently.
In the model of \citet{mulders_why_2021}, the presence of an iceline, which both cuts the pebble flux by half and reduces the pebble sizes, is crucial in delaying the inner disc growth and allowing pebble filtering to be efficient around FGK stars.
This is in general agreement with our results that show that some additional mechanism of inner growth delay needs to be employed in order to have efficient pebble filtering from outer giants. Such a mechanism could take the form of an iceline (cfr. Fig.\,\ref{fig:iceline_eff}) or efficient viscous heating (cfr. Fig.\,\ref{fig:multiple_filter_nofilter}).
Alternatively, \citet{chachan_small_2023} argued that M stars are simply more efficient at converting pebbles into planetary cores at short orbital periods, leading to the observed ``excess'' of super-Earths around M stars. The accretion efficiency around low-mass stars is higher due to their lower disc aspect ratio and slower pebble drift rate. 
A positive conditional occurrence rate is then the result of more massive stars being born with with more massive discs, which can nucleate both outer giants and super-Earths, leading to a higher conditional occurrence rate of super-Earths and cold giants.

\subsection{Terrestrial planet formation around the Sun}

In this study we do not aim to directly investigate terrestrial planet formation around our Sun. However, given the presence of Jupiter in the Solar System, and the lack of super-Earths, our model argues that the solar protoplanetary disc experienced a certain degree of viscous heating (cfr. Fig.\,\ref{fig:param_explor}).

The precise disc conditions, setting the pebble accretion efficiency, may thus regulate how rocky planet formation proceeds. 
Planetary growth may be completed within the disc phase through predominantly pebble accretion \citep{Levison_growing_terr_2015,Johansen_2021}. However, it could also be less efficient and therefore stalled at Mars-mass embryos, similar to the outcomes shown in Fig.\,\ref{fig:param_explor}, that then later undergo mutual collisions post disc dissipation \citep{lambrechts_formation_2019}.
Or, when pebble accretion is suppressed, it could be driven dominantly through planetesimal accretion \citep{morbidelli_contemporary_2022,batygin_formation_2023}.

It is clear that the mechanism underlying the formation of terrestrial planets in our Solar System remains elusive, with numerous models only partially capable of reproducing this process to varying degrees. Given the promising results from both pebble and planetesimal accretion models, a combined approach appears to be a reasonable direction for future studies to capture the complexities of planet formation in the Solar System.

\section{Summary and conclusions}
\label{sec:conclusion}
In this work, we constructed a 1D semi-analytical model for a pebble-driven planet formation scenario.
The initial planetary embryos' masses were taken from the top of the streaming instability mass distribution.
We then performed simulations of systems of multiple planets in order to investigate the role of mutual pebble filtering among planets on preventing super-Earth formation in the inner disc. For this reason, we considered a nominal pebble flux, $F_0$, large enough to create a Jupiter-like planet within the disc lifetime. 
Our simulations explored three possible disc models, one purely dominated by stellar irradiation (\texttt{mod:irrad}), and two accretion-heated discs, one moderately accretion-heated (\texttt{mod:surfheat}) and one strongly accretion-heated (\texttt{mod:midheat}). 

We started by investigating the role of the fragmentation velocity in determining the dominant pebble size, which also sets the pebble accretion efficiency, and thus the timescale of accretion of each planet. 
Pebble sizes are generally drift limited in the outer disc, but become fragmentation limited as they drift inward. This transition occurs at wider radii in case of low fragmentation velocities: around $50$ AU for $1 \: \rm m/s$, while it is located inside $1$\, AU for higher $v_{\rm frag}$ near $10 \: \rm m/s$  (see Fig.\,\ref{fig:st_frag}). Therefore, inside the water iceline, pebble sizes become strongly dependent on the assumed fragmentation velocity. The disc model choice also affects the dominant pebble size, with strong viscous heating leading to increased midplane temperatures, that significantly limit the size of fragmentation-limited pebbles, potentially leading to order of magnitude size differences.
Higher fragmentation velocities also lead to a faster transition between the 3D and 2D Hill accretion regimes, shortening the planet formation timescale, especially in the inner disc, regardless of the disc model (see Fig.\,\ref{fig:irr_vfrag}).

We find that accretion heating (i.e. \texttt{mod:surfheat} and \texttt{mod:midheat}) significantly slows down the pebble accretion timescale in the inner disc, especially in the case of low fragmentation velocities ($1 \: \rm m/s$). 
For moderate heating (\texttt{mod:surfheat}), the delay in the growth timescale mainly affects planets within 1 AU, 
promoting growth at orbital radii between 1 and 15 AU. 
This breaks the typical inside-out growth seen in a purely irradiated disc (cfr. Fig.\,\ref{fig:t_visc}).
The growth delay in moderately accretion-heated discs (\texttt{mod:surfheat}) does not prevent the formation of super-Earths in the inner disc by itself (middle panel of Fig.\,\ref{fig:param_explor}).
However, it can succeed in delaying and possibly halting the growth of the innermost embryo population, if combined with mutual pebble filtering due to the simultaneous formation of outer giants (middle panel of Fig.\,\ref{fig:param_explor}).
In the strong midplane heating case (\texttt{mod:midheat}), we observed that planet formation in the inner disc is largely suppressed, with growth timescales to the pebble isolation mass surpassing typical disc lifetimes (right panel, Fig.\,\ref{fig:t_visc}). This, combined with mutual filtering, is enough to suppress super-Earth formation in most cases (right panel of Figs. \ref{fig:multiple_filter_nofilter} and \ref{fig:param_explor}), even if dust leaking through the planetary gap is allowed (bottom row, Fig.\,\ref{fig:param_explor_50_0}).

The effect of mutual filtering due to the growth of an outer giant planet varies drastically based on the disc model. In the irradiated case it is basically irrelevant because of the inside-out growth mode: inner embryos grow to isolation mass faster than outer ones, which only filter a negligible amount of material before the inner embryos reach isolation mass (left panel of Fig.\,\ref{fig:multiple_filter_nofilter}).
By using a viscously heated disc, the delay in inner embryo's growth helps the effect of mutual filtering. In moderately heated disc, the effect is still small enough to prevent only the growth of the innermost embryos (around 0.1 AU, cfr. central panel of Fig.\,\ref{fig:param_explor}). 
In case of a strongly heated disc, the inner embryos growth is efficiently suppressed, out to the position of a Jupiter-like planet
(see Fig.\,\ref{fig:multiple_filter_nofilter} and Fig.\,\ref{fig:param_explor}).

The role of the iceline is less clear and model-choice dependent. 
In an irradiated disc, cutting the pebble flux by $50 \%$ at the iceline and considering mutual filtering among planets is still not enough to prevent inner embryos from reaching isolation mass (left panel, Fig.\,\ref{fig:iceline_eff}).
By combining the flux reduction with an increase in either the vertical turbulence or fragmentation alpha inside the iceline, the growth in the inner disc is hindered even in the irradiated case. However, why precisely such conditions capable of preventing super-Earth formation would be present in some discs, and not others, is unclear.

We also explored the possibility of pebbles leaking through the outer giant planet's gap once it reaches isolation mass. In a surface-heated disc, a $50 \%$ pebble leaking efficiency is sufficient to promote super-Earth formation in the inner disc (central top panel of Fig.\,\ref{fig:param_explor_50_0}). However, in the midplane-heated disc model, while the leaking slightly enhances planetary growth outside $1$ AU, it still fails to promote super-Earth formation (bottom central panel of Fig.\,\ref{fig:param_explor_50_0}).

To summarise, we show that pebble filtering due to outer giants is generally not efficient enough to prevent super-Earth formation, unless some mechanism of delaying inner disc growth is also simultaneously employed. These could be, for example, efficient viscous heating of the inner disc or the presence of an ice line which significantly reduces the pebble flux and particle size.
However, if super-Earths dominantly form through pebble accretion, such conditions must be rare, given that observational trends argue for only a mild suppression, if any, of super-Earth occurrence in systems with cold giants. 

\begin{acknowledgements}
CD thanks Lizxandra Flores-Rivera and Adrien Houge for helpful discussions.
ML acknowledges the ERC starting grant 101041466-EXODOSS.
The authors thank the anonymous referee for the report that helped to improve the quality of this manuscript.
\end{acknowledgements}

\bibliographystyle{aa} 
\bibliography{bibliography.bib}

\begin{appendix}

\section{Accretion heating}
\label{app:visc_heating}
\subsection{Vertical temperature profile}
\label{subsec:Tz}

In the viscously heated inner part of the disc, the midplane temperature can be expressed as a function of the deposited accretion heat. 
Below the photosphere of the disc ($z_{\rm ph}$), the radiative flux $F$ takes the form
\begin{align}
    F(z) = - \frac{4}{3} \frac{\sigma_{\rm SB}}{ \kappa \rho}  \frac{dT^4(z)}{dz},
\end{align}
assuming an optically thick region in radiative equilibrium \citep{Hubeny_1990}. 
Here, $\sigma_{\rm SB}$ is the Stefan-Boltzmann constant and $\kappa$ is the Rosseland mean opacity.
Assuming a z-independent effective dust opacity ($\kappa_{\rm eff}$), we can simplify the integral
\begin{align}
    T^4(z) - T_{\rm ph}^4  
   = \frac{3}{4} \frac{\kappa_{\rm eff}}{\sigma_{\rm SB} } 
   \int_{z}^{z_{\rm ph}}  \rho(z) F(z) dz\,.
\end{align} 
We consider accretion heating to take place dominantly in an arbitrarily thin layer located at a height $z_Q<z_{ph}$. A standard assumption is midplane viscous heating, $z_Q=0$ \citep{Hubeny_1990, Menou2004}, while non-ideal magneto-hydrodynamic simulations have shown that accretion heating takes place several gas scale heights above the disc \citep{Mori_2019,Bethune_2020}. 
In this latter case, the radiative flux cancels out in the midplane layer between the symmetric upper and lower heating layers, leading to a constant midplane temperature $T_{\rm mid} \approx T(z<z_{\rm q})$, given by 
\begin{align}
    T_{\rm mid}^4  
   = \frac{3}{4} \frac{\kappa_{\rm eff}}{\sigma_{\rm SB} } 
   \int_{z_Q}^{z_{\rm ph}}  \rho(z) F(z) dz + T_{\rm ph}^4 \,.
\end{align}
We can further simplify the calculation by considering a z-independent deposition of the full accretion heating in the considered heating layer, obtaining
\begin{align}
    F(z) = \frac{1}{2} Q_+ = \frac{1}{2} \int_{-\infty}^{+\infty} q(z) dz \,.
\end{align}
This allows us to express the midplane temperature as 
\begin{align}
\label{eq:T_Mid}
    T_{\rm mid}^4  
   &= \frac{3}{2^3} \frac{Q_+ }{\sigma_{\rm SB} } \kappa_{\rm eff}
   \int_{z_Q}^{z_{\rm ph}} \rho(z)dz + T_{\rm ph}^4 \,.
\end{align}
Finally, we introduce an efficiency parameter $\epsilon_{\rm el}$ expressing the degree by which the heating layer is elevated above the midplane
\begin{align}
    \int_{z_Q}^{z_{\rm ph}} \rho(z)dz  
    &= \frac{\Sigma_{\rm gas}}{2} \frac{2}{\sqrt{2\pi}}\int_{z_Q/H}^{\infty} \exp\left({-\frac{1}{2} x^2} \right)dx \\
    &= \epsilon_{\rm el} \frac{\Sigma_{\rm gas}}{2} \,.
\end{align}
This last integral is the complementary error function. We get
$\epsilon_{\rm el} = 1$ for $z_Q=0$, and numerically
$\epsilon_{\rm el} = 4.6\cdot 10^{-2} $ for $z_Q=2H$, and
$\epsilon_{\rm el} = 6.3\cdot 10^{-5}$ for $z_Q=4H$.

Inside the terrestrial planet-forming region, the heating layer is approximately located between $2$ to $4$ scale heights above the midplane \citep{Mori_2021,Kondo_2023}. Therefore, we will conservatively assume $\epsilon_{\rm el}=10^{-2}$ as our nominal value.

Inserting the efficiency parameter $\epsilon_{\rm el}$ in Eq.\,\ref{eq:T_Mid} the midplane temperature becomes
\begin{align}
    T_{\rm mid}^4  
   &= \frac{3}{2^4} \epsilon_{\rm el} \frac{Q_+ }{\sigma_{\rm SB} } \kappa_{\rm eff} \Sigma_{\rm gas} + T_{\rm ph}^4 \,.
   \label{eq:Tmidgen}
\end{align}
In the limit where all the accretion heating is deposited in the midplane ($\epsilon_{\rm el} \approx 1 $), we obtain the standard expression for viscous heating \citep[][ ignoring the Planck mean absorption term]{Hubeny_1990,Menou2004}. 
In contrast, when midplane heating is inefficient ($\epsilon_{\rm el} \ll 1 $) , it is possible to approach the lower midplane temperature limit where $T_{\rm mid} \approx T_{\rm ph}$.

\subsection{Midplane temperature as function of accretion rate}

\subsubsection{Standard steady-state accretion disc}
In a standard steady-state accretion disc \citep{Pringle_1981}, the accretion rate onto the host star ($\dot{M}_{\star}$) can be linked to the disc viscosity $\nu$ as 
\begin{align}
    \dot M_{\star} &= 3 \pi \Sigma_{\rm gas} \nu \,.
    \label{eq:steadystate}
\end{align}
When expressing this term with an alpha prescription, $\nu = \alpha_{\nu} c_s^2 \Omega_{\rm K}^{-1}$ \citep{Shakura1973}, with $c_s$ the midplane sound velocity, we obtain the following standard expression for the gas surface density
\begin{align}
    \Sigma_{\rm gas} 
    &= \frac{\dot M_{\star}}{3 \pi \nu}  \\
    &= \frac{1}{3\pi} \frac{1}{\alpha_{\nu}}  \frac{\Omega_{\rm K}}{c_s^2} \dot M_{\star} \,.
    \label{eq:SigMdot}
\end{align}
Here, we consider $\alpha_{\nu}$ to be an effective alpha that drives accretion, which in reality may be driven in a complex way through active layers above the midplane and disc wind torques. Therefore, it expresses the link between the effective gas surface density and accretion rate onto the host star, rather than being a measure of midplane turbulence. 
We take as a nominal value $\alpha_{\nu}=10^{-2}$ in agreement with disc lifetimes of several Myr \citep{Haisch2001}.

\subsubsection{Effective temperature as function of accretion rate}
The heating rate per unit surface, as function of the gas acccretion rate, can be expressed as 
\begin{align}
    Q_{+} =  \frac{9}{4} \Omega_{\rm K}^2  \frac{\epsilon_{\rm heat}\dot M_{\star}}{3\pi} \,, \label{eq:Qgeneral}
\end{align}
following \citet{Pringle_1981}.
This expression is chosen so it reduces to the classical prescription for maximally efficient shear midplane heating when $\epsilon_{\rm heat}$ is set to 1
\begin{align}
    Q_{+} &=  \frac{9}{4} \Omega_{\rm K}^2 \epsilon_{\rm heat} \Sigma_{\rm gas} \nu = \Sigma_{\rm gas} \nu \left( r \frac{d\Omega_{\rm K}}{dr} \right)^2 \,,
\end{align}
while also allowing for a lower fraction $\epsilon_{\rm heat}$ of the total accretion stress to be deposited as a heating source for the disc.

Using the balance between the heating rate and the cooling rate of the two disc faces
\begin{align}
    Q_+ = Q_- &= 2 \sigma_{\rm SB}T_{\rm eff}^4 \,,
\end{align}
we can express the effective temperature as
\begin{align}
    T_{\rm eff}^4 
    &=  \frac{Q_+}{2 \sigma_{\rm SB}} \\
    &=  \frac{3}{2^3 \pi}  \epsilon_{\rm heat} \frac{\Omega_{\rm K}^2 \dot{M}_{\star}}{\sigma_{\rm SB}} \,.
\end{align}
In what follows, we will assume for simplicity that $\epsilon_{\rm heat} =0.5$, which appears to be an appropriate upper bound \citep[][Appendix B]{Mori_2019,Mori_2021}.

\subsubsection{Midplane temperature as function of acccretion rate}
We can insert Eq.\,\ref{eq:Qgeneral} and Eq.\,\ref{eq:SigMdot} into Eq.\,\ref{eq:Tmidgen}, obtaining an expression for the midplane temperature as a function of the gas accretion rate
\begin{align}
    T_{\rm mid}^4  
    &\approx 
    \frac{3}{2^4} \epsilon_{\rm el} \frac{\kappa_{\rm eff}}{\sigma_{\rm SB}}
    \left( \frac{9}{4} \Omega_{\rm K}^2 \epsilon_{\rm heat} \frac{\dot M_{\star}}{3\pi}\right)
    \left( \frac{1}{3\pi} \frac{1}{\alpha_{\nu}}  \frac{\Omega_{\rm K}}{c_{\rm s}^2} \dot M_{\star} \right) \\
    &= \frac{3}{2^6 \pi^2}
     \epsilon_{\rm el} \epsilon_{\rm heat}\alpha_{\nu}^{-1}
     \frac{\kappa_{\rm eff}}{\sigma_{\rm SB}}
      c_{\rm s}^{-2} \Omega_{\rm K}^3 \dot M_{\star}^2 \,.\label{eq:Tmidapprox}
\end{align}
Here we have assumed in the fist step that $T_{\rm mid}>T_{\rm ph}=T_{\rm eff}$. As expected, the midplane temperature increases with opacity and accretion rate, but decreases moderately with increased $\alpha_{\nu}$ as this effectively lowers the local gas surface density.
Finally, note that $c_{\rm s}$ and, potentially, $\kappa_{\rm eff}$ are temperature dependent quantities.
Therefore, further progress requires specifying a (temperature-dependent) dust opacity. 

\subsection{Dust opacity}
\label{subsec:opa}
Determining the effective opacity is challenging. We assume the dominant opacity source to be small dust grains, as commonly done in the literature \citep[e.g.][]{Birnstiel_2018}. However, the presence of these particles may be reduced due to dust growth and settling in evolved discs \citep{Kondo_2023}. Here, we mainly aim to make a minimal physical model, leaving a more in depth exploration to future studies. 

The wavelength-dependent dust opacity, with an effective cross section $\sigma_{\rm gr}$, is given by 
\begin{align}
    \kappa_\lambda 
    &= Q_{\rm e}(x) \pi a_{\rm gr}^2 \frac{n_{\rm gr}}{\rho_{\rm gas}} \\
    &= Q_{\rm e}(x)  \frac{3}{4 a_{\rm gr} \rho_{\rm gr}} Z\,,
\end{align}
with $a_{\rm gr}$ and $\rho_{\rm gr}$ the dust grain size and density, respectively. The factor $Z$ is the dust-to-gas ratio of the grains that are the dominant opacity carrier.
The parameter $Q_{\rm e}(x)$ gives the extinction in terms of a size parameter
\begin{align}
    x = 2\pi \frac{a_{\rm gr}}{\lambda}\,.
\end{align}
This size parameter can be related to the disc temperature via Wien's displacement law, 
\begin{align}
    \nu \approx 2 \frac{k_{\rm B}}{h}T \,,
\end{align}
which gives 
\begin{align}
    x &\approx 16
      \left( \frac{a_{\rm gr}}{0.1\,\rm mm} \right)
      \left( \frac{T}{200\,\rm K} \right) \,.
\end{align}

We can then explore the limits of the extinction coefficient that can be summarised as 
\begin{align}
    Q_{\rm e}(x) \approx 
    \begin{cases}
       x & \text{ if } x < 1 \\
       2 & \text{ if }  x > 1
    \end{cases} \,.
\end{align}
A more in depth description, including grain porosity, can be found in \citet{Kataoka_2014}. 
The analysis by \citet{Savvidou_2020} shows that the $x>1$ regime broadly holds within the water ice line for generic size distributions with a maximum size up to $\approx 0.1$\,mm. 
We therefore only consider the limit $x>1$, which removes the wavelength dependency of the opacity. In this way, Planck opacities ($\kappa_{\rm P}$) and Rosseland opacities ($\kappa_{\rm R}$) are equal and we get 
\begin{align}
 \kappa_{\rm eff} = \kappa_{\rm P,x>1} = \kappa_{\rm R,x>1} 
 = \frac{3}{2 a_{\rm gr} \rho_{\rm gr}} Z\,,
\end{align}
which is independent of the temperature and inversely proportional to the grain size. 
For nominal grain values, we find
\begin{align}
  \kappa_{\rm eff} = 1.5\,
  \left( \frac{a_{\rm gr}}{0.1\,\rm mm} \right)^{-1}
  \left( \frac{\rho_{\rm gr}}{1\,{\rm g/cm}^3} \right)^{-1}
  \left( \frac{Z}{0.01} \right) {\rm cm}^2{\rm/g}\,.
\end{align}

\subsection{Inner disc midplane temperature (for x > 1)}

We can now compute, through Eq.\,\ref{eq:Tmidapprox}, the midplane temperature in the $x>1$ case
\begin{align}
    T_{\rm mid}^5 &\approx
    \frac{3^2}{2^7 \pi^2}
    \epsilon_{\rm el} \epsilon_{\rm heat} \alpha_{\nu}^{-1}
    \frac{\mu m_{\rm H}}{\sigma_{\rm SB} k_{\rm B}}
    \frac{Z}{a_{\rm gr} \rho_{\rm gr}}
    \Omega_{\rm K}^3 \dot{M}_{\star}^2 \label{eq:Tmid5}
\end{align}
where we made use of $c_{\rm s}^2 = (k_{\rm B}/(\mu k_{\rm B}))T$, with $\mu=2.34$.
A common derived quantity from the midplane temperature is the gas disc aspect ratio
\begin{align}
   \frac{H}{r} = \frac{c_{\rm s}}{v_{\rm K}} \,.
\end{align}
Making use of Eq.\,\ref{eq:Tmid5}, the gas disc aspect ratio takes the form 
\begin{align}
    \left( \frac{H}{r}\right)^{10} 
    &=
    \frac{3^2}{2^7 \pi^2}
    \frac{\epsilon_{\rm el} \epsilon_{\rm heat}}{ \alpha_{\nu}}
    \frac{1}{\sigma_{\rm SB}}
    \left( \frac{\mu m_{\rm H}}{ k_{\rm B}} \right)^{-4}
    \frac{Z}{a_{\rm gr} \rho_{\rm gr}}
    (GM_\star)^{-7/2} r^{1/2}
    \dot M_\star^2 .
\end{align}
Normalising the expression to our nominal values we can write
\begin{align}
 \left( \frac{H}{r} \right)_{\rm visc} \approx&\,
     0.019
     \left( \frac{\epsilon_{\rm el}}{10^{-2}} \right)^{1/10}
     \left( \frac{\epsilon_{\rm heat}}{0.5} \right)^{1/10}
     \left( \frac{\alpha_{\nu}}{10^{-2}} \right)^{-1/10}
     \notag\\
     &\left(\frac{Z}{0.01} \right)^{1/10}
     \left( \frac{a_{\rm gr}}{0.1\,{\rm mm}} \right)^{-1/10}
     \left(\frac{\rho_{\rm gr}}{1\,{\rm g\,cm}^3} \right)^{-1/10}   
     \notag\\
     &\left( \frac{r}{\rm AU} \right)^{1/20}
     \left( \frac{\dot M_\star}{10^{-8} M_\odot{\rm yr}^{-1}} \right)^{1/5}
     \left( \frac{M_\star}{M_\odot}\right)^{-7/20}
     \,.
     \label{eq:Hrvisc}
\end{align}
This expression illustrates that the viscously heated inner-disc scale height is only weakly dependent on orbital radius ($\propto r^{1/20}$). However, it strongly depends on the gas accretion rate onto the central star, which evolves by orders of magnitude: from young discs accreting at rates above 
$\dot M_\star = 10^{-7} M_\odot{\rm yr}^{-1}$, to less than 
$10^{-9} M_\odot{\rm yr}^{-1}$ closer to disc dissipation (see Eq.\ref{eq:sigma-gas_dot}). 
The gas scale height depends weakly on the choice of $\alpha_{\nu}$, $\epsilon_{\rm heat}$, and $\epsilon_{\rm el}$. 
However, $\epsilon_{\rm el}$ differs by orders of magnitude depending on the height of heating layer (Section \ref{subsec:Tz}), therefore acting as a key variable in setting the gas scale height. 
The properties of the opacity-carrying grains $Z$, $a_{\rm gr}$, and $\rho_{\rm gr}$ only weakly influence the inner disc aspect ratio, but the opacity contribution is certainly model dependent (Section \ref{subsec:opa}).

This result can be compared with other expressions for midplane viscously heated gas scale heights. When setting $\epsilon_{\rm heat}=1$ and $\epsilon_{\rm el}=1$, Eq.\,\ref{eq:Hrvisc} reduces to 
\begin{align}
  \frac{H}{r} \approx& \:
     0.041 
     \left( \frac{\alpha_{\nu}}{10^{-3}} \right)^{-1/10}
     \left( \frac{r}{\rm AU} \right)^{1/20}
     \left( \frac{\dot M_\star}{10^{-8} M_\odot{\rm yr}^{-1}} \right)^{1/5} \,,
     \label{eq:Hrvisc2}
\end{align}
which is here normalised to $\alpha_{\nu}=10^{-3}$ to facilitate the comparison to the frequently used fitting formula by \citet{ida_radial_2016}, based on the work by \citet{Oka_2011}. 
Conveniently, we recover the same power law scalings on orbital radius $r$, accretion rate $\dot M_\star$ and $\alpha_{\nu}$. 
However, at 1 AU, \citet{ida_radial_2016} reports a scale height of $H/r = 0.027$ for $\dot M_\star = 10^{-8} M_\odot{\rm yr}^{-1}$ and $\alpha_{\nu}=10^{-3}$, which is below what we find here (see Fig.\,\ref{fig:disc_quantities}).
In contrast, the expression in \citet[][his Appendix B]{liu_super-earth_2019} is comparable, finding a scale height of $H/r = 0.045$ at $1$ AU for $\dot M_\star = 10^{-8} M_\odot{\rm yr}^{-1}$ and $\alpha_{\nu}=10^{-3}$. Their aspect ratio weakly decreases with increasing orbital radius. This is due to their choice of a dust opacity that is linearly dependent on temperature.

\section{Drift limited Stokes number}
\label{app:st_drift}
In the protoplanetary disc, as dust coagulates and grows bigger, it decouples from the gas and feels a headwind that causes it to inward-drift. This happens when the growth timescale of the dust gets comparable to the drift timescale, leading to a limit dimension that a certain dust grain can reach at a given position in the disc before inevitably drifting inwards: the drift limit. 
The drift-limited Stokes number can be computed by equating the drift timescale (\ref{eq:t_drift}) with the growth timescale (\ref{eq:t_gr}).
The drift timescale is the ratio between the position in the disc and the radial velocity of the dust grain:
\begin{equation}
    \label{eq:t_drift}
    t_{\mathrm{drift}} = \frac{r}{v_r},
\end{equation}
where $v_r$ is the pebble drift velocity given by \citep{nakagawa_settling_1986, guillot_filtering_2014,ida_radial_2016}:
\begin{equation}
    \label{eq:v_r}
    v_r = - 2 \frac{\mathrm{St}}{1+\mathrm{St^2}} \eta v_{\mathrm{K}} + \frac{\mathrm{1}}{1+\mathrm{St^2}} \eta v_{\mathrm{K}} u_{\nu},
\end{equation}
where $u_{\nu} \approx -\nu/r \approx - \alpha (H/r)^2 v_{\mathrm{K}}$ is the radial viscous diffusion velocity. 
Here we approximate $v_r \approx -2\mathrm{St} \eta v_{\mathrm{K}}$, considering that typically $\mathrm{St} \ll 1$, thus $\mathrm{St^2}+1 \approx 1$ and the second term in equation (\ref{eq:v_r}) is mostly negligible.
The growth timescale is expressed as the ratio between the dimension of the grain and its growth rate
\begin{equation}
\label{eq:t_gr}
    t_{\mathrm{growth}} = \frac{a_{\rm gr}}{\Dot{a}_{\rm gr}}. 
\end{equation}
The grain growth rate follows \citep{birnstiel_simple_2012}:
\begin{equation}
    \Dot{a}_{\rm gr} = \frac{\rho_{\mathrm{dust}} \Delta v}{\rho_{\rm gr}},
\end{equation}
where $\rho_{\mathrm{dust}}$ is the dust volume density, $\rho_{\rm gr}$ is the grain density and $\Delta v$ expresses the relative velocity between the grains, which in case of turbulent motion and similar-sized dust grains is given by \citep{ormel_closed-form_2007}
\begin{equation}
    \label{eq:delta_v}
    \Delta v = \frac{1}{4} \sqrt{3 \alpha_z \mathrm{St}} c_{\mathrm{s}},
\end{equation}
where $\alpha_z$ is the vertical stirring of the grains.
Using the definition of midplane density both for the dust and the gas $\rho = \Sigma/(\sqrt{2 \pi} H)$ and approximating $H_{\mathrm{dust}} \approx \sqrt{\alpha_z/\mathrm{St}} H_{\mathrm{gas}}$ we can rewrite the grain growth rate as
\begin{equation}
    \label{eq:a_dot}
    \dot{a}_{\rm gr} = \frac{\sqrt{3}}{4} \frac{\rho_{\mathrm{gas}}}{\rho_{\mathrm{\rm gr}}} \mathrm{St} c_{\mathrm{s}} \frac{\Sigma_{\rm dust}}{\Sigma_{\rm gas}}
\end{equation}
Using Eqs. (\ref{eq:t_gr}) and (\ref{eq:a_dot}), the growth timescale can be written as
\begin{equation}
        t_{\mathrm{growth}} = a_{\rm gr} \frac{4}{\sqrt{3}} \frac{\rho_{\mathrm{\rm gr}}}{\rho_{\mathrm{gas}}} \frac{1}{\mathrm{St} \: c_{\mathrm{s}}} \frac{\Sigma_{\rm gas}}{\Sigma_{\rm dust}}
\end{equation} 
where the only assumption made so far is that the relative velocity between grains is due to turbulent motion (see Eq.\,\ref{eq:delta_v}).
The expression that relates the dimension of the grain $a_{\rm gr}$ with its Stokes number depends on the drag regime that the grains are subjected to: Epstein drag (if $a_{\rm gr} < 9/4 \: \lambda_{\mathrm{mfp}}$) or Stokes drag (if $a_{\rm gr} > 9/4 \: \lambda_{\mathrm{mfp}}$). In the first case the Stokes number is linear in the grain dimension, in the latter it depends quadratically on the grain dimension \citep{ida_radial_2016}
\begin{equation}
\label{eq:st_ep_reg_app}
\begin{cases}
   \mathrm{St_{Ep}} = \frac{\rho_{\rm gr} \Omega_{\mathrm{K}}}{\rho_{\mathrm{gas}} c_{\mathrm{s}}} a_{\rm gr} & a_{\rm gr} <\frac{9}{4} \lambda_{\mathrm{mfp}}, \\
   \mathrm{St_{St}} = \frac{4}{9}\frac{\rho_{\rm gr}}{\rho_{\mathrm{gas}}}\frac{\Omega_{\mathrm{K}}}{c_{\mathrm{s}} \lambda_{\mathrm{mfp}}} a_{\rm gr}^2 & a_{\rm gr} >\frac{9}{4} \lambda_{\mathrm{mfp}}, 
\end{cases}
\end{equation}
where $\lambda_{\mathrm{mfp}}$ the gas molecule mean free path.
At this point we can equate the growth and the drift timescales in the two respective drag regimes to obtain an explicit expression for the Stokes number of the particle.
\paragraph{Epstein regime}
We can compute the growth timescale in Epstein regime by using the particle size in Eq. (\ref{eq:st_ep_reg}) 
\begin{equation}
    \label{eq:t_g_E}
    t_{\mathrm{growth, Ep}} = \frac{4}{\sqrt{3} \Omega_{\mathrm{K}}\epsilon_{\rm{p}}} \frac{\Sigma_{\rm gas}}{\Sigma_{\rm dust}},
\end{equation}
where we introduced the parameter $\epsilon_{\mathrm{p}} = 0.5$ that represents the coagulation efficiency between grains. We notice that the growth timescale is independent on the density of the grain.
Using the continuity equation $F_{\mathrm{peb}} = 2 \pi r v_r \Sigma_{\mathrm{dust}}$ we can express the pebble surface density through the flux and then equate growth time scale (\ref{eq:t_g_E}) with drift time scale (\ref{eq:t_drift}) to get the pebble Stokes number in the Epstein drag regime
\begin{equation}
\label{eq:st_d_ep}
    \mathrm{St_{drift, Epstein}} = \sqrt{\frac{\sqrt{3} \epsilon_{\mathrm{p}}F_{\mathrm{peb}}}{32 \pi \Sigma_{\mathrm{gas}}  \eta^2 v_{\mathrm{K}}r}},
\end{equation}
where we assumed $\mathrm{St^2+1 \approx 1}$ and the pebble flux $F_{\mathrm{peb}}$ to be Stokes independent.

\paragraph{Stokes regime}
When the dimension of the pebbles becomes comparable to the gas mean free path, they enter the Stokes drag regime.
In this case, the growth timescale will be given by
\begin{equation}
    \label{eq:t_g_S}
    t_{\mathrm{growth, St}} = \frac{3}{\sqrt{3}} \frac{\Sigma_{\rm gas}}{\Sigma_{\rm dust}} \epsilon_{\mathrm{p}}^{-1} \sqrt{\frac{\lambda_{\mathrm{mfp}} \: \rho_{\rm gr}}{\Omega_{\mathrm{K}}  \mathrm{St_{\mathrm{St}}} c_{\mathrm{s}} \: \rho_{\mathrm{gas}}}},
\end{equation}
where we also added the coagulation efficiency coefficient $\epsilon_{\mathrm{p}}$.
Again, making use of the continuity equation and equating growth and drift timescale we can calculate the explicit form of the Stokes number in Stokes drag regime
\begin{equation}
    \label{eq:st_drift_st_app}
    \mathrm{St_{drift, Stokes}} = \left(\frac{48\pi}{\sqrt{3}} \frac{\Sigma_{\mathrm{gas}}}{F_{\mathrm{peb}}} \sqrt{\frac{\rho_{\rm gr} \lambda_{\mathrm{mfp}}}{\rho_{\mathrm{gas}}H}} \frac{ \eta^2 v_{\mathrm{k}}^2}{\epsilon_{\mathrm{p}}\Omega_{\mathrm{K}}}\right)^{-2/3},
\end{equation}
where $\lambda_{\mathrm{mfp}} = (\mu m_{\mathrm{p}})/(\sigma_{\mathrm{H}} \rho_{\mathrm{gas}})$ is the gas mean free path.
Summarising, the drift-limited Stokes number is given by equation (\ref{eq:st_d_ep}) if the pebble is drifting in Epstein regime and by equation (\ref{eq:st_drift_st_app}) if it is subjected to a Stokes regime.

\begin{figure*}
    \centering
    \includegraphics[width=\linewidth]{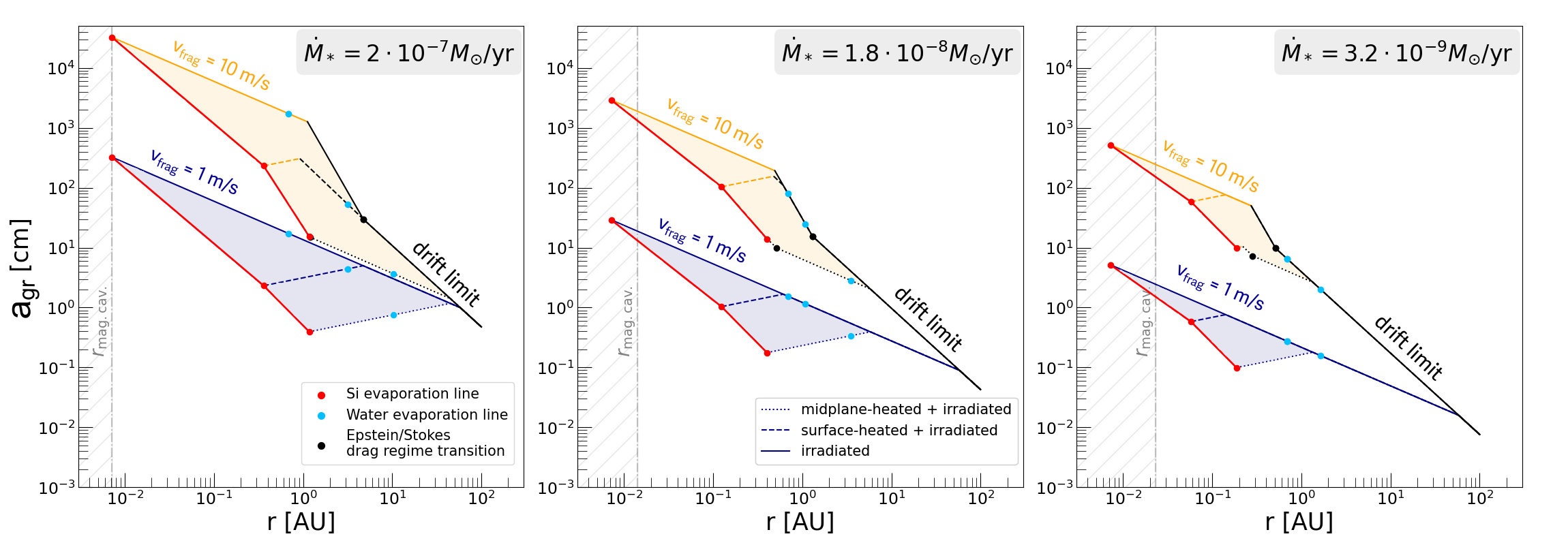}
    \caption{Pebble dimension as a function of their position for three different gas accretion rates (corresponding to $t \simeq 0.1$ Myr (left), $t \simeq 1.0$ Myr (central), and $t \simeq 5.0$ Myr (right)
), different fragmentation velocities (blue and orange branches) and different disc models (solid, dashed and dotted lines). This plot is complementary to Fig.\,\ref{fig:st_frag} but shows dimensions in cm instead of Stokes numbers.}
    \label{fig:R_peb_r}
\end{figure*}

By means of Eqs. (\ref{eq:st_ep_reg_app}), one can derive the pebble dimension (in cm) given its Stokes number. This is shown in Fig.\,\ref{fig:R_peb_r}, which represents the same pebbles as Fig.\,\ref{fig:st_frag} but translated into their actual grain sizes rather than Stokes numbers.

\end{appendix}

\end{document}